\documentclass[aps,prc,twocolumn,superscriptaddress,showpacs,floatfix]{revtex4-1}
\usepackage{graphicx,amssymb}
\usepackage[fleqn]{amsmath}
\usepackage{amsfonts}
\usepackage{dcolumn}
\usepackage{bm}
\usepackage{braket}
\usepackage{multirow}
\usepackage{MnSymbol}

\newcommand{\intt}
{
	\displaystyle \int
}

\newcommand{\hatvec}[1]
{
	\hat{ \vec{#1} }
}

\newcommand{\diffn}[3]
{
	\frac{ { \partial }^{ #3 } #1 }{ \partial { #2 }^{ #3 } }
}

\newcommand{\unitsText}[1]
{
	\, \text{#1}
}

\begin{document}

\preprint{ }

\title{Gamow shell model description of radiative capture reactions $^6$Li$(p,\gamma)$$^7$Be and $^6$Li$(n,\gamma)$$^7$Li }

\author{G.X. Dong}
\affiliation{Grand Acc\'el\'erateur National d'Ions Lourds (GANIL), CEA/DSM - CNRS/IN2P3,
BP 55027, F-14076 Caen Cedex, France}

\author{N. Michel}
\affiliation{Grand Acc\'el\'erateur National d'Ions Lourds (GANIL), CEA/DSM - CNRS/IN2P3,
BP 55027, F-14076 Caen Cedex, France}

\author{K. Fossez}
\affiliation{NSCL/FRIB Laboratory,
Michigan State University, East Lansing, Michigan  48824, USA}

\author{M. P{\l}oszajczak}
\affiliation{Grand Acc\'el\'erateur National d'Ions Lourds (GANIL), CEA/DSM - CNRS/IN2P3,
BP 55027, F-14076 Caen Cedex, France}

\author{Y. Jaganathen}
\affiliation{NSCL/FRIB Laboratory,
Michigan State University, East Lansing, Michigan  48824, USA}

\author{R.M. Id Betan}
\affiliation{Physics Institute of Rosario (CONICET), Bv. 27 de Febrero 210 bis, S2000EZP Rosario, Argentina}
\affiliation{Department of Physics and Chemistry FCEIA(UNR), Av. Pellegrini 250, S2000BTP Rosario, Argentina}

\date{\today}

\begin{abstract}
\noindent
  {{\bf Background:}\\ According to standard stellar evolution, lithium abundance is believed to
   be a useful indicator of the stellar age. However, many evolved stars like red giants show huge fluctuations around expected theoretical abundances that are not yet fully understood. The better knowledge of nuclear reactions that contribute to the creation and destruction of lithium can help to solve this puzzle. \\
  {\bf Purpose:} In this work we apply the Gamow shell model (GSM) formulated in the coupled-channel representation (GSM-CC) to investigate the mirror radiative capture reactions $^6$Li$(p,\gamma)$$^7$Be and $^6$Li$(n,\gamma)$$^7$Li.  \\
  {\bf Method:} GSM offers the most general treatment of couplings between discrete resonant states and the non-resonant continuum. The cross-sections are calculated using a translationally invariant Hamiltonian with the finite-range interaction which is adjusted to reproduce spectra, binding energies and one-nucleon separation energies in $^{6-7}$Li, $^7$Be. The reaction channels are built by coupling the wave functions of ground state $1_1^+$ and excited states $3^+_1$, $0^+_1$, $2^+_1$ of $^6$Li with the projectile wave function in different partial waves. \\
  {\bf Results:}  We include all relevant $E1$, $M1$, and $E2$ transitions from the initial continuum states to the final bound states $J={3/2}_1^-$ and $J={1/2}^-$ of $^7$Li and $^7$Be.
   Our microscopic astrophysical factor for the $^6$Li(p,$\gamma$)$^7$Be reaction follows the average trend of the experimental value as a function of the center of mass energy. For ${ {}^{6}\text{Li} ( n , \gamma ) {}^{7}\text{Li} }$, the calculated cross section agrees well with the data from the direct measurement of this reaction at stellar energies.
 \\
  {\bf Conclusion:} We demonstrate that the $s$-wave radiative capture of proton (neutron) to the first excited state $J^{\pi}=1/2_1^+$ of $^7$Be ($^7$Li) is crucial and increases the total astrophysical $S$-factor by about 40 \%.
}
  
\end{abstract}


\maketitle
\section{Introduction}
\label{intro}
The structure of low-lying states in nuclei around the beta-stability valley, i.e., low-energy spectra, nuclear moments, electromagnetic transitions, is well described by the standard shell model (SM). In a vicinity of the neutron or proton drip lines, the atomic nucleus becomes weakly bound or even unbound in the ground state, and hence the description of its basic properties requires an explicit consideration of the coupling to the scattering continuum and decay channels. The comprehensive description of bound states, resonances and scattering many-body states within a single theoretical framework is possible in the 
Gamow shell model (GSM)~\cite{Michel02,Michel03,Michel09}. The attempts to reconcile the SM with the reaction theory inspired the development of the continuum shell model~\cite{rf:7,rf:9} and the GSM in the channel representation \cite{Jaganathen14,Fossez15}.

GSM is the rigged Hilbert space generalization of the SM~\cite{Michel09}. Many-body states are expanded in the basis of Slater determinants spanned by bound, resonance and (complex-energy) non-resonant scattering states of the complete single particle (s.p.) Berggren ensemble~\cite{rf:4}. In the past, GSM has been applied mainly to describe the structural properties of many-body bound states, resonances and their decays. However, due to the lack of separation between different decay channels in GSM, one can not describe the nuclear reaction processes directly. In order to describe both the nuclear structure and reactions in a unified theoretical framework, the GSM has been recently formulated in the coupled-channel (CC) representation~\cite{Jaganathen14,Fossez15}. The GSM-CC approach has been applied to the low-energy elastic and inelastic proton scattering~\cite{Jaganathen14} and proton or neutron radiative capture reactions~\cite{Fossez15}.

The low-energy proton radiative capture reactions play an important role in the nuclear astrophysics, in particular in the nucleosynthesis of light and medium-heavy elements. In recent years, much interest has been devoted to the study of reactions which can produce $^7$Be in the stellar environment~\cite{Adelberger11}, especially to the $^6$Li$(p,\gamma)$$^7$Be reaction which is crucial for the consumption of $^6$Li and the formation of $^7$Be. This reaction can contribute to the understanding of solar neutrino problem and pp-II, pp-III reaction chains since it produces $^7$Be which is destroyed by the $^7$Be$(p,\gamma)$$^8$B reaction. The $^6$Li$(p,\gamma)$$^7$Be reaction has been studied either by direct proton capture measurements~\cite{Bashkin55,Ostojic83,Switkowski79,Cecil92,He13} or by the analyzing-power experiments with the polarized proton beams~\cite{Prior04}. Theoretical studies of this reaction have been done in the potential model~\cite{Barker80,Angulo99,Camargo08,Huang10,Xu13} and in various cluster model approaches~\cite{Arai02,Dubovichenko11,Nesterov11}.

$^6$Li$(n,\gamma)$$^7$Li is the mirror reaction of $^6$Li$(p,\gamma)$$^7$Be. In nuclear astrophysics, lithium isotopes have attracted a great interest because of the puzzled abundance of $^6$Li and $^7$Li. Whereas $^7$Li in hot, low-metallicity stars is supposed to come from the Big Bang nucleosynthesis, $^6$Li is believed to originate from the spallation and fusion reactions in the interstellar medium~\cite{Jedamzik00}. Therefore, the abundance ratio of $^6$Li and $^7$Li could be considered as an effective time scale of the stellar evolution~\cite{Vangioni-Flam99}. In this context, the $^6$Li$(n,\gamma)$$^7$Li reaction is a direct bridge between these two isotopes whose cross section may influence their abundance ratio significantly. The $^6$Li$(n,\gamma)$$^7$Li cross section at stellar energies either has been measured by the direct neutron capture~\cite{Ohsaki00}, or extracted from the inverse reaction $^7$Li$(\gamma,n)$$^6$Li~\cite{Karataglisis89,Bramblett73,Green64}. Theoretical investigations of this reaction have been done using either the direct capture model~\cite{Su10} or the cluster approach~\cite{Dubovichenko13-1,Dubovichenko13-2}.

In the present paper, we will apply the microscopic GSM-CC approach to study the low-energy spectra and cross-sections in the mirror radiative capture reactions: $^6$Li$(p,\gamma)$$^7$Be and $^6$Li$(n,\gamma)$$^7$Li. The paper is organized as follows. The general formalism of GSM-CC approach and the solution of the GSM-CC equations are briefly presented in Sec.~\ref{sec-2}. 
Results of GSM-CC calculations are discussed in Sec.~\ref{sec3}. The detailed descriptions of low-energy spectra of $^7$Be and $^7$Li, as well as the proton and neutron radiative capture processes on the target of $^6$Li are given in Sec.~\ref{sec3a} and Sec.~\ref{sec3b} respectively. Finally, the main conclusions of the work are summarized in Sec.~\ref{sec4}.

\section{The Gamow shell model in the coupled-channel representation}
\label{sec-2}

\subsection{The GSM Hamiltonian}
\label{sec-2a}

To remove spurious center of mass (c.m.) excitations in the GSM wave functions, the Hamiltonian is usually written in the intrinsic nucleon-core coordinates of the cluster-orbital shell model (COSM)~\cite{Ikeda88}:
\begin{equation}
	\hat{H} = \sum_{i = 1}^{ {N}_{ \text{val} } } \left( \frac{ \hat{\vec{p}}_{i}^{2} }{ 2 { \mu }_{i} } + {U}_{\text{core}} ( \hat{r}_{i} ) \right) + \sum_{i < j}^{ {N}_{ \text{val} } } \left( V ( \hat{\vec{r}}_{i} - \hat{\vec{r}}_{j} ) + \frac{ {\hat{\vec{p}}_{i}}{\cdot} {\hat{\vec{p}}_{j} }}{ {M}_{\text{core}} } \right)
	\label{GSM_Hamiltonian}
\end{equation}
where $N_\text{val}$ is the number of valence nucleons, $M_{\text{core}}$ is the mass of the core, $\mu_i$ is the reduced proton or neutron mass, $U_{\text{core}}(\hat{r})$ is the single-particle (s.p.) potential which describes the field of a core acting on each nucleon. The last term in Eq.~(\ref{GSM_Hamiltonian}) represents the recoil term, and $V(\hat{\vec{r}}_i-\hat{\vec{r}}_j)$ is the two-body interaction between valence nucleons.

By introducing a one-body mean-field ${ U ( \hat{r}_{i} ) }$, the GSM Hamiltonian can be recast in a form:
\begin{equation}
{ \hat{H} = \hat{U}_{ \text{basis} } + \hat{T} + \hat{V}_{ \text{res} } }
\label{eq_H_GSM}
\end{equation}
where ${ \hat{T} }$ is the kinetic term, ${ \hat{U}_{ \text{basis} } }$ is the potential which generates the s.p. basis:
\begin{equation}
	\hat{U}_{ \text{basis} } = \sum_{ i = 1 }^{ {N}_{ \text{val} } } ( {U}_{\text{core}} ( \hat{r}_{i} ) + U ( \hat{r}_{i} ) )
	\label{eq_U_basis}
\end{equation}
and the residual interaction is given by ${ \hat{V}_{ \text{res} }}$:
\begin{equation}
	\hat{V}_{ \text{res} } = \sum_{i < j}^{ {N}_{ \text{val} } } \left( V ( \hatvec{r}_{i} - \hatvec{r}_{j} ) + \frac{ {\hatvec{p}_{i}}{\cdot} {\hatvec{p}_{j} }}{ {M}_{\text{core}} } \right) - \sum_{ i = 1 }^{ {N}_{ \text{val} } } U ( \hat{r}_{i} ).
	\label{eq_H_res}
\end{equation}

\subsection{The GSM coupled-channel equations}
\label{sec-2b}
The antisymmetric eigenstates of GSM-CC can be expanded in the complete basis of channel states ( ${ \ket{ r , c } = \hat{ \mathcal{A} } ( \ket{r} \otimes \ket{c}  }$):
\begin{eqnarray}
&&\ket{ c } = \ket { T_c ; \ell_c ~ j_c ~ \tau_c} \label{channel_definition} \\
&&\hat{ \mathcal{A} }\ket{ \Psi } = \ket{ \Psi } = \sumint\limits_{c} \intt_{0}^{ \infty } dr \, {r}^{2} \frac{ {u}_{c} (r) }{r} \ket{ r , c } \label{eq_expan_eig_full_channel_basis}
\end{eqnarray}
where $\ket{T_c}$ is the target state, $\ell_c$, $j_c$ and $\tau_c$ are the orbital momentum, 
total momentum and isospin projection quantum numbers of the projectile, respectively,
${ \hat{ \mathcal{A} } }$ is the antisymmetrization operator, and
${u}_{c} (r) /r$ are the antisymmetrized channel wave functions.

By inserting Eq.~\eqref{eq_expan_eig_full_channel_basis} in the Schr\"odinger equation and then projecting it onto a given channel basis state ${ \bra{ r' , c' } }$, one obtains the GSM-CC equations:
\begin{equation}
	\sumint\limits_{c} \intt_{0}^{ \infty } dr \, {r}^{2} \left( {H}_{ c' , c } ( r' , r ) - E {O}_{ c' , c } ( r' , r ) \right) \frac{ {u}_{c} (r) }{r} = 0,
	\label{eq_CC_eqs_general}
\end{equation}
where
\begin{equation}
	{H}_{ c' , c } ( r' , r ) = \braket{ r' , c' | \hat{H} | r , c }
	\label{eq_CC_H_ME_general}
\end{equation}
and
\begin{equation}
	{O}_{ c' , c } ( r' , r ) = \braket{ r' , c' | r , c }
	\label{eq_CC_O_ME_general}
\end{equation}
are the Hamiltonian and the norm matrix elements in the channel representation, respectively.

The channel state $\ket{ r , c }$ can be constructed using a complete Berggren set of s.p. states~\cite{rf:4} which includes bound states, resonances, non-resonant scattering states from the contour in the complex $k$-plane \cite{Michel02,Michel03,Michel09}:
\begin{equation}
	\ket{ r , c } = \sum_{i} \frac{ {u}_{i} (r) }{r} \ket{ { \phi }_{i}^{ \text{rad} } , c }
	\label{eq_CC_basis_state_expansion_Berggren_2}
\end{equation}
where ${ \ket{ { \phi }_{i}^{ \text{rad} } , c } = \hat{ \mathcal{A} } ( \ket{ { \phi }_{i}^{ \text{rad} } } \otimes \ket{c} ) }$, ${u}_{i} (r) / r  = { \braket{ { \phi }_{i}^{ \text{rad} } | r }}$, and $\ket{ { \phi }_{i}^{ \text{rad} }}$ is the radial part.

For large projectile momentum, the antisymmetry between the low-energy target states and the high-energy projectile states can be neglected. Hence, the expansion
~\eqref{eq_CC_basis_state_expansion_Berggren_2} can be written as:
\begin{align}
	\ket{ r , c } &= \sum_{ i = 1 }^{ {i}_{ \text{max} } - 1 } \frac{ {u}_{i} (r) }{r} \ket{ { \phi }_{i}^{ \text{rad} } , c } \nonumber \\
	&+ \ket{r} \otimes \ket{c} - \sum_{ i = 1 }^{ {i}_{ \text{max} } - 1 } \frac{ {u}_{i} (r) }{r} \ket{ { \phi }_{i;c_{\rm proj}} } \otimes \ket{ {c}_{ \text{targ} } }
	\label{xxx}
\end{align}
where ${ {i}_{ \text{max} } }$ denotes the index from which the antisymmetry effects are neglected, $\ket{ { \phi }_{i;c_{\rm proj}} }$ and $\ket{ {c}_{ \text{targ} }}$ represent the projectile and target states, $\ket{r} \otimes \ket{c}$ and $\ket{ { \phi }_{i;c_{\rm proj}} } \otimes \ket{ {c}_{ \text{targ} }}$ stand for non-antisymmetrized states. In the case ${ i \geq {i}_{ \text{max} } }$, the GSM Hamiltonian \eqref{eq_H_GSM} splits into ${ \hat{H}_{ \text{proj} } }$ and ${ \hat{H}_{ \text{targ} } }$ terms
acting on ${ \ket{ { \phi }_{i;c_{\rm proj}} } }$ and ${ \ket{ {c}_{ \text{targ} } } }$, respectively:
\begin{align}
	& \hat{H}_{ \text{proj} } \ket{ { \phi }_{i;c_{\rm proj}} } = {E}_{ i , {c}_{ \text{proj} } } \ket{ { \phi }_{i;c_{\rm proj}} }; \nonumber \\
	& \hat{H}_{ \text{targ} } \ket{ {c}_{ \text{targ} } } = {E}_{ {c}_{ \text{targ} } } \ket{ {c}_{ \text{targ} } } \ .
	\label{eq_CC_E_proj}
\end{align}

Then, the matrix elements of the Hamiltonian ${ {H}_{ c' , c } ( r' , r ) }$ and the overlap ${ {O}_{ c' , c } ( r' , r ) }$ can be calculated with the help of expansion~\eqref{xxx} and Eq.~\eqref{eq_CC_E_proj}. One obtains:
\begin{align}
	{H}_{ c' , c } ( r' , r ) &= - \frac{ { \hbar }^{2} }{ 2 \mu_c } \left( \frac{1}{r} \diffn{ ( r \cdot ) }{r}{2} - \frac{ \ell_c ( \ell_c + 1 ) }{ {r}^{2} } - {k}_{ {c}_{ \text{targ} } }^{2} \right) \nonumber \\
	&\times \frac{ \delta ( r - r' ) }{ {r}^{2} } { \delta }_{ c' , c } + {V}_{ c' , c } ( r' , r ) \ .
	\label{eq_CC_EM_H_final}
\end{align}
In this equation, ${k}_{ {c}_{ \text{targ} } }^{2} = 2 \mu {E}_{ {c}_{ \text{targ} } } /{ { \hbar }^{2} }$ and $\mu_c$ is the reduced mass of the projectile in the channel $c$. The channel-channel coupling potential ${V}_{ c' , c } ( r' , r )$ is given by:
\begin{equation}
	{V}_{ c' , c } ( r' , r ) = {U}_{ \text{basis} } (r) \frac{ \delta ( r - r' ) }{ {r}^{2} } { \delta }_{ c' , c } + \tilde{V}_{ c' , c } ( r' , r )
	\label{eq_CC_ME_Vccp_pot}
\end{equation}
where $\tilde{V}_{ c' , c } ( r' , r )$ contains the channel couplings and exchange terms of the Hamiltonian:
\begin{align}
	&\tilde{V}_{ c' , c } ( r' , r ) = \sum_{i,i' = 1 }^{{i}_{ \text{max} }} \frac{ {u}_{ i' } ( r' ) }{ r' } \frac{ {u}_{i} (r) }{r}  \braket{ i' , c' | \hat{H} | i , c } \nonumber \\
	&\quad - \sum_{ i = 1 }^{ {i}_{ \text{max} } - 1 } \frac{ {u}_{ i } ( r' ) }{ r' } \frac{ {u}_{i} (r) }{r} ( {E}_{ i , {c}_{ \text{proj} } }  + {E}_{ {c}_{ \text{targ} } } ) { \delta }_{ c' , c }\ ,
	\label{eq_CC_EM_Vccp_pot_rest}
\end{align}
where $\ket{ i , c } = \hat{ \mathcal{A} } ( \ket{i} \otimes \ket{c})$.
In the same way, for ${ {O}_{ c' , c } ( r' , r ) }$ one obtains:
\begin{equation}
	{O}_{ c' , c } ( r' , r ) = \frac{ \delta ( r - r' ) }{ {r}^{2} } { \delta }_{ c' , c } + \tilde{O}_{ c' , c } ( r' , r )
	\label{eq_CC_EM_O_final}
\end{equation}
where $\tilde{O}_{ c' , c } ( r' , r )$ contains the channel couplings and exchange terms arising from the antisymmetrization of channels:
\begin{align}
	\tilde{O}_{ c' , c } ( r' , r ) &= \sum_{ i,i' = 1 }^{{i}_{ \text{max} }} \frac{ {u}_{ i' } ( r' ) }{ r' } \frac{ {u}_{i} (r) }{r} \braket{ i' , c' | O | i , c } \nonumber \\
	&- \sum_{ i = 1 }^{ {i}_{ \text{max} } - 1 } \frac{ {u}_{ i } ( r' ) }{ r' } \frac{ {u}_{i} (r) }{r} { \delta }_{ c' , c } \ .
	\label{eq_CC_EM_Occp_rest}
\end{align}

Due to the antisymmetry between the projectile and target states, different channel basis states $\ket{ r , c }$ are nonorthogonal which leads to a generalized eigenvalue problem. To solve it, the orthogonal channel basis states ($\ket{ r , c }_{o} = \hat{O}^{ -\frac{1}{2} } \ket{ r , c } \label{eq_CC_non_ortho_to_ortho_channel_states}$) are applied:
\begin{equation}
	{}_{o}\braket{ r' , c' | r , c }_{o} = \frac{ \delta ( r' - r ) }{ {r}^{2} } { \delta }_{ c' c } 
	\label{eq_CC_ortho_channel_basis_braket}
\end{equation}
where ${ \hat{O} }$ is the overlap operator. Then, the GSM-CC equations~\eqref{eq_CC_eqs_general} become:
\begin{align}
	\sumint\limits_{c} \intt_{0}^{ \infty } dr \, {r}^{2} &( {}_{o}\braket{ r' , c' | \hat{H}_{o} | r , c }_{o} - E {}_{o}\braket{ r' , c' | \hat{O} | r , c }_{o} ) \nonumber \\
	&\times {}_{o}\braket{ r , c | { \Psi }_{o} } = 0 
	\label{eq_CC_eqs_general_clear_ortho}
\end{align}
where
${ \hat{H}_{o} = \hat{O}^{  \frac{1}{2} } \hat{H} \hat{O}^{  \frac{1}{2} } }$, and $ {\ket{ \Psi_o } = \hat{O}^{1/2} \ket{ \Psi } }$. With a substitution: ${ \ket{ \Phi } = \hat{O} \ket{ \Psi } }$, the generalized eigenvalue problem in Eq.~\eqref{eq_CC_eqs_general_clear_ortho} can be transformed into a standard one:
\begin{equation}
	\sumint\limits_{c} \intt_{0}^{ \infty } dr \, {r}^{2} ( {}_{o}\braket{ r' , c' | \hat{H} | r , c }_{o} - E {}_{o}\braket{ r' , c' | r , c }_{o} ) {}_{o}\braket{ r , c | \Phi } = 0 \ .
	\label{eq_CC_eqs_general_clear_ortho_again}
\end{equation}
In the nonorthogonal channel basis, these CC equations become:
\begin{equation}
	\sumint\limits_{c} \intt_{0}^{ \infty } dr \, {r}^{2} \braket{ r' , c' | \hat{H}_{m} | r , c } \frac{ {w}_{c} (r) }{r} = E \frac{ {w}_{ c' } ( r' ) }{ r' } 
	\label{eq_CC_final}
\end{equation}
where ${w}_{c} (r)/ r \equiv \braket{ r , c | \hat{O}^{ \frac{1}{2} } | \Psi } = {}_{o}\braket{ r , c | \Phi }$, and ${ \hat{H}_{m} = \hat{O}^{ - \frac{1}{2} } \hat{H} \hat{O}^{ - \frac{1}{2} } }$ is the modified Hamiltonian.

In the calculation of the matrix elements of ${ \hat{H}_{m} }$, it is convenient to introduce a new operator ${ \hat{ \Delta } }$ ($\hat{O}^{ - \frac{1}{2} } = \hat{ \Delta } + \hat{\textbf 1}\label{eq_CC_Delta_def}$). Then, the matrix elements of ${ \hat{H}_{m} }$ are calculated as:
\begin{equation}
	{H}_{m} = ( \Delta + \hat{\textbf 1} ) H ( \Delta + \hat{\textbf 1} ) = H + H \Delta + \Delta H + \Delta H \Delta \ .
	\label{eq_CC_H_m_trick}
\end{equation}
By inserting \eqref{eq_CC_H_m_trick} in CC equations \eqref{eq_CC_final} and replacing matrix elements ${ \braket{ r' , c' | \hat{H} | r , c } }$ using Eqs.~\eqref{eq_CC_EM_H_final}-\eqref{eq_CC_ME_Vccp_pot}, one obtains the CC equations for the reduced radial wave functions ${ {w}_{c} (r) / r }$:
\begin{align}
		& \left( - \frac{ { \hbar }^{2} }{ 2 \mu_c } \left( \frac{1}{r} \diffn{ ( r \cdot ) }{r}{2} - \frac{ \ell_c ( \ell_c + 1 ) }{ {r}^{2} } \right) + {V}_{c}^{ ( \text{loc} ) } (r) \right) \frac{ {w}_{c} (r) }{r}  \nonumber \\
&+ \sum_{ c' } \intt_{0}^{ \infty } dr' \, r { r' }^{2} \frac{ {V}_{ c , c' }^{ ( \text{non-loc} ) } ( r , r' ) }{ r r' } \frac{ {w}_{ c' } ( r' ) }{ r' } \nonumber \\
		&= ( E - {E}_{ {c}_{ \text{targ} } } ) \frac{ {w}_{c} (r) }{r}  \ .
		\label{eq_CC_modif_Eqs_full_2}
	\end{align}
In this equation,  ${ {V}_{c}^{ ( \text{loc} ) }}$ is the local potential: ${ {V}_{c}^{ ( \text{loc} ) } (r) = {U}_{ \text{basis} } (r) }$, and the non-local potential is given by:
\begin{align}
	\frac{1}{ r' r } {V}_{ c' , c }^{ ( \text{non-loc} ) } &( r' , r ) = \tilde{V}_{ c' , c } ( r' , r ) + \braket{ r' , c' | \hat{H} \hat{ \Delta } | r , c } \nonumber \\
	&+ \braket{ r' , c' | \hat{ \Delta } \hat{H} | r , c } + \braket{ r' , c' | \hat{ \Delta } \hat{H} \hat{ \Delta } | r , c } \ .
	\label{eq_CC_non_local_pot_final}
\end{align}
The radial channel wave functions ${ {u}_{c} (r) / r }$ are then obtained from the solutions of Eq. \eqref{eq_CC_modif_Eqs_full_2} using the equation:
\begin{equation}
	\frac{ {u}_{c} (r) }{r} = \frac{ {w}_{c} (r) }{r} + \sum_{ c' } \intt_{0}^{ \infty } dr' \, { r' }^{2} \braket{ r , c | \hat{O}^{ \frac{1}{2} } \hat{ \Delta } \hat{O}^{ \frac{1}{2} } | r' , c' } \frac{ {w}_{ c' } ( r' ) }{ r' } \ .
	\label{eq_uc_wc_link}
\end{equation}

\subsection{Solution of the GSM-CC equations}
\label{sec-2c}
Using a generalization of the equivalent-potential method \cite{Fossez15}, the local ${V}_{c}^{ ( \text{loc})}(r)$ and non-local ${V}_{ c , c' }^{ ( \text{non-loc} ) } ( r , r' )$ potentials in Eq. \eqref{eq_CC_modif_Eqs_full_2} are replaced by the equivalent local potential ${V}_{ c , c' }^{ ( \text{eq} ) } (r)$: 
\begin{align}
	&{V}_{ c , c' }^{ ( \text{eq} ) } (r) = {V}_{c}^{ ( \text{loc} ) } (r) { \delta }_{ c' , c } \nonumber \\
	&\quad + \sum_{ c' } \intt_{0}^{ \infty } dr' \, \frac{ 1 - {F}_{ c' } (r) }{ {w}_{ c' } (r) } {V}_{ c , c' }^{ ( \text{non-loc} ) } ( r , r' ) {w}_{ c' } ( r' ) \ ,
	\label{eq_CC_equi_pot_def}
\end{align}
and the source term ${S}_{c} (r)$:
\begin{equation}
	{S}_{c} (r) =  \sum_{ c' } \intt_{0}^{ \infty } dr' \, {F}_{ c' } (r) {V}_{ c , c' }^{ ( \text{non-loc} ) } ( r , r' ) {w}_{ c' } ( r' ) \ .
	\label{eq_CC_source_def}
\end{equation}
${ {F}_{c'} (r) }$ in Eqs. \eqref{eq_CC_equi_pot_def}, \eqref{eq_CC_source_def} is the smoothing function \cite{Fossez15} to cancel divergences of the equivalent potential ${ {V}_{ c , c' }^{ ( \text{eq} ) } (r) }$ close to the zeroes of ${ {w}_{c} (r) }$.

With these substitutions, the GSM-CC equations \eqref{eq_CC_modif_Eqs_full_2} become
\begin{align}
	\diffn{ {w}_{c} (r) }{r}{2} &= \left( \frac{ \ell_c ( \ell_c + 1 ) }{ {r}^{2} } - {k}_{c}^{2} \right) {w}_{c} (r) \nonumber \\
	&+ \frac{ 2 \mu_c }{ { \hbar }^{2} } \left( \sum_{ c' } {V}_{ c , c' }^{ ( \text{eq , sy} ) } (r) {w}_{ c' } (r) + {S}_{c}^{ ( \text{sy} ) } (r) \right)
	\label{eq_CC_modif_Eqs_full_6}
\end{align}
where
${k}_{c}^{2} = { 2 \mu_c } ( E - {E}_{ {c}_{ \text{targ} } } )/{ { \hbar }^{2} }
\label{eq_CC_kc_def}$.
Eqs. \eqref{eq_CC_modif_Eqs_full_6} are solved iteratively to determine the equivalent potential, the source term, and the mutually orthogonal radial wave functions ${ {w}_{c} (r) }$. Starting point for solving these equations is provided by a set of radial channel wave functions ${ \{ {w}_{c} (r) \} }$ obtained by the diagonalization of GSM-CC equations \eqref{eq_CC_final} in the Berggren basis of channels.

\section{Discussion and results}
\label{sec3}

The differential cross-sections for proton or neutron radiative capture reactions can be expressed using  the matrix elements of electromagnetic operators between the antisymmetrized initial and final states. The detailed discussion of relevant cross-sections and various approximations in these calculations, such as the long-wavelength approximation and the treatment of antisymmetry in the many-body matrix elements of the electromagnetic operators, can be found in Ref.~\cite{Fossez15}. 

The radiative capture cross section for a final state of the total angular momentum ${ {J}_{f} }$ is:
\begin{equation}
	{ \sigma }_{ {J}_{f} } ( {E}_{ \text{c.m.} } ) = \intt_{0}^{ 2 \pi } d{ \varphi }_{ \gamma } \intt_{0}^{ \pi } \sin{ \theta }_{ \gamma } d{ \theta }_{ \gamma } \frac{ d{ \sigma }_{ {J}_{f} } ( {E}_{ \text{c.m.} } , { \theta }_{ \gamma } , { \varphi }_{ \gamma } ) }{ d{ \Omega }{ \gamma } }
	\label{eq_rad_cap_partial_cross_section}
\end{equation}
and the total cross section is thus:
\begin{equation}
	\sigma ( {E}_{ \text{c.m.} } ) = \sum_{ {J}_{f} } { \sigma }_{ {J}_{f} } ( {E}_{ \text{c.m.} } ) \ .
	\label{eq_rad_cap_total_cross_section}
\end{equation}
Instead of the total cross-section, for the radiative capture of charged particles one often uses the astrophysical $S$ factor:
\begin{equation}
	S ( {E}_{ \text{c.m.} } ) = \sigma ( {E}_{ \text{c.m.} } ) {E}_{ \text{c.m.} } {e}^{ 2 \pi \eta }
	\label{eq_rad_cap_astrophysical_factor}
\end{equation}
which removes the exponential dependence of the cross section at low energies due to the Coulomb barrier. ${ \eta }$ in Eq. \eqref{eq_rad_cap_astrophysical_factor} is the Sommerfeld parameter: $\eta = (mZ_1Z_2)/(\hbar^2k)$, where $Z_1$ and $Z_2$ are the proton numbers of the projectile and target nuclei.

\subsection{Results for the $^{6}\text{Li}(p,\gamma)^{7}\text{Be}$ reaction}
\label{sec3a}

In the GSM and GSM-CC calculations for $^6$Li, $^7$Be, and $^{6}\text{Li}(p,\gamma)^{7}\text{Be}$ we take $^4$He as the inert core. For each considered partial wave: $l$=0, 1, and 2, the potential generated by the core is given by the Woods-Saxon (WS) potential with the spin-orbit term (see Table~\ref{para-1}) and the Coulomb potential of radius $r_{\text{Coul}}$=2.5 fm. For the two-body force, we use the Furutani-Horiuchi-Tamagaki (FHT) finite-range two-body interaction~\cite{Furutani78,Furutani79}. Parameters of the FHT interaction, which were adjusted to reproduce binding energies of low-lying states and proton separation energies in $^6$Li and $^7$Be, are given in Table~\ref{para-2}.
Detailed introduction of the FHT interaction in the context of GSM calculation can be found in Ref.~\cite{Fossez15}. 

GSM and GSM-CC calculations are performed in two resonant shells: $0p_{3/2}$ and $0p_{1/2}$, and several shells in the non-resonant continuum along the discretized contours: $\mathcal{L}^+_{s_{1/2}}$, $\mathcal{L}^+_{p_{1/2}}$, $\mathcal{L}^+_{p_{3/2}}$, $\mathcal{L}^+_{d_{3/2}}$, and $\mathcal{L}^+_{d_{5/2}}$. Each contour consists of three segments joining the points: $k_{\text{min}}$=0.0, $k_{\text{peak}}=0.15-i0.14$ fm$^{-1}$, $k_{\text{middle}}$=0.3 fm$^{-1}$ and $k_{\text{max}}$=2.0 fm$^{-1}$, and each segment is discretized with 7 points.
Hence, GSM and GSM-CC calculations are performed in 107 shells: 22 $p_{3/2}$ and $p_{1/2}$ shells, and 21 $s_{1/2}$, $d_{3/2}$ and $d_{5/2}$ shells. To reduce the size of the GSM Hamiltonian matrix, the basis of Slater determinants is truncated by limiting the occupation of $p_{3/2}$, $p_{1/2}$, $s_{1/2}$, $d_{3/2}$ and $d_{5/2}$ scattering states to two particles.

\begin{table}[ht!]
\caption{Parameters of the WS potential of the $^4$He core used in the GSM and GSM-CC description of $^6$Li, $^7$Be, and $^{6}\text{Li}(p,\gamma)^{7}\text{Be}$.\label{para-1}}
\begin{ruledtabular}
\begin{tabular}{ccc}
  Parameter & Protons & Neutrons \\
  \hline
  $a$ & 0.65 fm & 0.65 fm \\
  $R_0$ & 2.0 fm & 2.0 fm \\
  $V_\text{o}$ & 47.571 MeV & 52.212 MeV \\
  $V_\text{so}(l=1)$ & 6.14 MeV & 3.088 MeV \\
  $V_\text{so}(l=2)$ & 6.14 MeV & 3.088 MeV \\
\end{tabular}
\end{ruledtabular}
\end{table}

\begin{table}[ht!]
\caption{Parameters of the FHT interaction in GSM and GSM-CC calculations in $^6$Li, $^7$Be, and $^{6}\text{Li}(p,\gamma)^{7}\text{Be}$. The superscripts C, SO, and T stand for central, spin-orbit, and tensor, respectively, and the indices ``s" and ``t" stand for singlet and triplet.\label{para-2}}
\begin{ruledtabular}
\begin{tabular}{cc}
  Parameter & Value \\
    \hline
  $\nu^\textsf{C}_{\textsf{t,t}}$  & 13.883 MeV \\
  $\nu^\textsf{C}_{\textsf{s,t}}$  & -8.510 MeV \\
  $\nu^\textsf{C}_{\textsf{s,s}}$  & -14.158 MeV \\
  $\nu^\textsf{C}_{\textsf{t,s}}$  & -7.226 MeV \\
  $\nu^\textsf{SO}_{\textsf{t,t}}$  & -1181.084 MeV \\
  $\nu^\textsf{SO}_{\textsf{s,t}}$  & 0 MeV \\
  $\nu^\textsf{T}_{\textsf{t,t}}$  & 17.86 MeV fm$^{-2}$ \\
  $\nu^\textsf{T}_{\textsf{s,t}}$  & -1.298 MeV fm$^{-2}$ \\
\end{tabular}
\end{ruledtabular}
\end{table}

\subsubsection{Spectrum of $^7$Be}

In GSM calculations with the FHT interaction (see Table \ref{para-2}), the ground state of a target nucleus $^6$Li is bound with respect to $^4$He by 3.658 MeV, which is close to the experimental value 3.698 MeV. Reaction channels in GSM-CC calculations are built by coupling the ground state $J_{\text{targ}}^\pi=1^+$ and the excited states $J_{\text{targ}}^\pi=3^+_1, 0^+_1$ and $2^+_1$ of $^6$Li with the proton in partial waves: $s_{1/2}$, $p_{1/2}$, $p_{3/2}$, $d_{3/2}$ and $d_{5/2}$. The composite states of $^7$Be $([^6\text{Li}(J_{\text{targ}}^\pi)\otimes p(l,j)]^{J_f^\pi})$ are ${3/2}_1^-$, ${1/2}_1^-$ bound states, and ${7/2}_1^-$, ${5/2}_1^-$, ${5/2}_2^-$ resonances.

\begin{figure}[htb]
\includegraphics[width=0.9\linewidth]{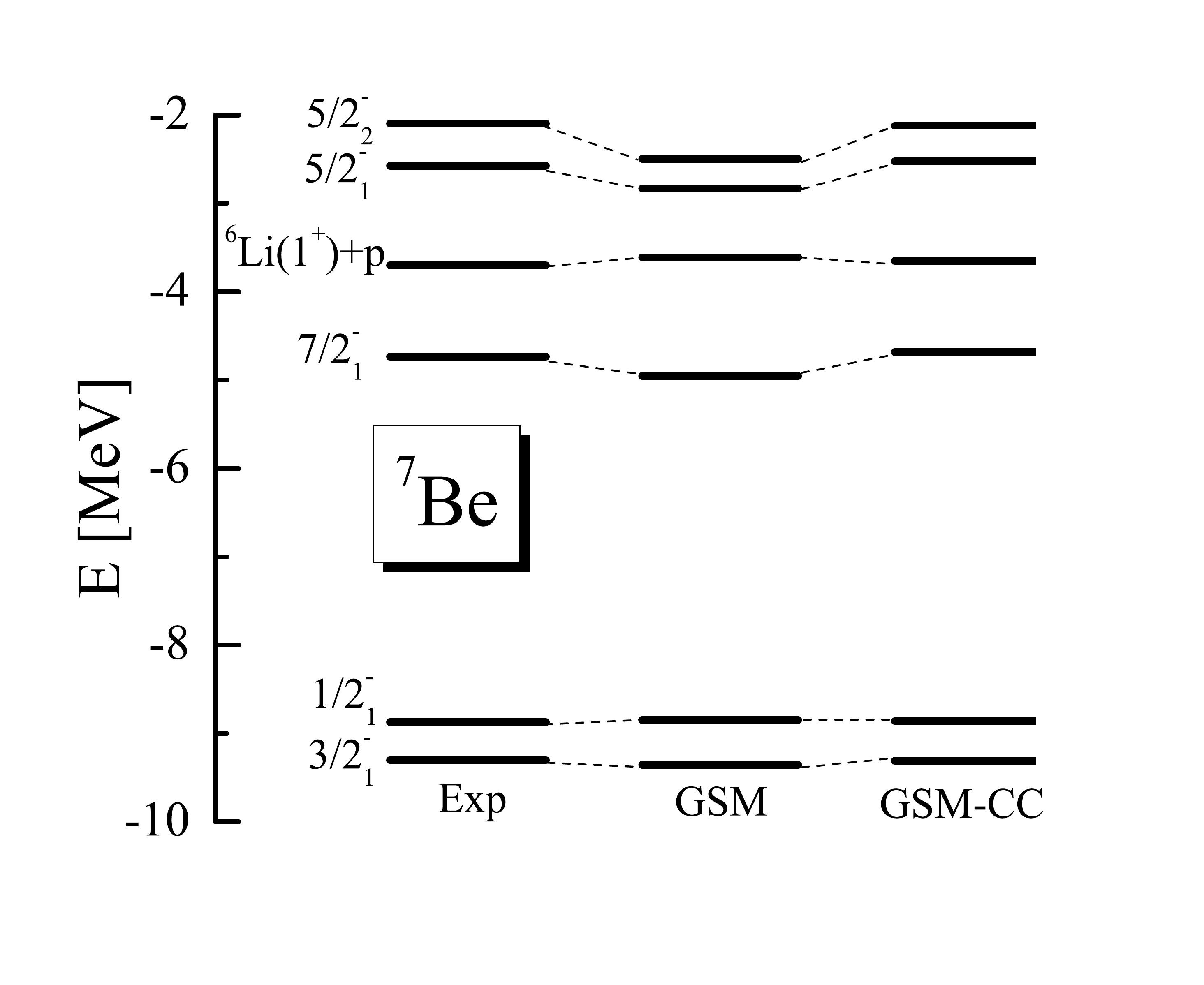}
\caption{Calculated energy levels of $^7$Be in GSM and GSM-CC are compared to the experimental data~\cite{nndc}. Energies are given relative to the energy of $^4$He.}
\label{fig-1}
\end{figure}

The energy spectrum and the proton separation energy in $^7$Be, as obtained in GSM and GSM-CC calculations, are compared with the experimental data~\cite{nndc} in Fig.~\ref{fig-1}. In GSM-CC calculations, the two-body part of ${\hat H}$ from which the channel-channel coupling potentials $V_{c,c'}$ are calculated, have been rescaled by the small multiplicative corrective factors $c(J^\pi)$ for ${J^\pi=3/2}_1^-$, ${1/2}_1^-$ states to compensate for the missing correlations due to the omission of non-resonant channels build by coupling the continuum states of $^6$Li with the proton in different partial waves. These scaling factors are: $c(\frac{3}{2}^-)=1.021$, $c(\frac{1}{2}^-)=1.051$.

One can see in Fig.~\ref{fig-1} that the calculated energy levels and proton separation energies are in good agreement with the experimental data, especially for the low-lying bound states ${3/2}^-_1$ and ${1/2}^-_1$. The ${7/2}^-_1$ state, which lies in-between the $[{^4}\text{He} - {^3}\text{He}]$ and  
$[{^6}\text{Li} - \text{p}]$ decay thresholds, is a narrow resonance. In our GSM-CC calculation, this state is bound because the $[^4\text{He}(J_{\rho}^\pi)\otimes ^3\text{He}(J_{\rho}^\pi)]^{J_f^\pi}$ reaction channel is missing in the basis.

The broad resonances ${5/2}^-_1$ ($\Gamma^{\rm exp}\simeq 1.2$ MeV) and ${5/2}^-_2$ ($\Gamma^{\rm exp}\simeq 0.4$ MeV), lying above the $[{^4}\text{He} - {^3}\text{He}]$ and  
$[{^6}\text{Li} - \text{p}]$ decay thresholds, are also resonances in our framework and decay by the proton emission. The calculated width is smaller than found experimentally, partially due to the missing $[^4\text{He}(J_{\rho}^\pi)\otimes ^3\text{He}(J_{\rho}^\pi)]^{J_f^\pi}$ reaction channel.

\subsubsection{Astrophysical $S$-factor for $^{6}\text{Li}(p,\gamma)^{7}\text{Be}$ reaction}

Having calculated the antisymmetrized initial ($^6$Li) and final ($^7$Be) GSM wave functions, we can begin the discussion of proton radiative capture cross-section calculation for the reaction $^{6}\text{Li}(p,\gamma)^{7}\text{Be}$.

The description of electromagnetic transitions requires effective charges for proton and neutron. We use the standard values for $E1$ and $E2$ effective charges.
For $E1$ transitions, we take~\cite{Hornyak75}:
\begin{equation}
	{e}_{ \text{eff} }^{p} = e \left( 1 - \frac{Z}{A} \right) \ ; \qquad
	{e}_{ \text{eff} }^{n} = - e \frac{Z}{A}
	\label{eq1}
\end{equation}
${ Z }$ and ${ A }$ are the proton number and the total number of nucleons, respectively.
The standard values for E2 transitions are
\begin{equation}
	{e}_{ \text{eff} }^{p} = e \left( 1 - \frac{Z}{A} + \frac{Z}{ {A}^{2} } \right) \ ; \qquad
	{e}_{ \text{eff} }^{n} = - e \frac{Z}{ {A}^{2} }.
	\label{eq2}
\end{equation}
There are no effective charges for $M1$ transitions.

Resonances in the spectrum of a composite ${ A }$-nucleon system ($^7$Be) correspond to the peaks in the radiative capture cross section at the c.m. energy:
${E}_{ \text{c.m.} } = {E}_{i}^{ ( A ) } [ \text{GSM-CC} ] - {E}_{0}^{ ( A - 1 ) } [ \text{GSM} ]
\label{eq_rad_cap_resonance_position}$.
Here ${ {E}_{i}^{ ( A ) } [ \text{GSM-CC} ] }$ is the GSM-CC energy of the resonance '${ i }$' in the nucleus ${ A }$, and ${ {E}_{0}^{ (A) } [ \text{GSM} ] }$ is the GSM ground state energy of the target nucleus ${ ( A - 1 ) }$ ($^6$Li in the studied case).

The experimental proton separation energy in ${ {}^{6}\text{Li} }$ is $S_{\rm p}^{(\rm exp)}=4.593$ MeV. The final nucleus ${ {}^{7}\text{Be} }$ has three states: ${3/2}_{1}^{-}$, ${1/2}_{1}^{-}$, and ${7/2}_{1}^{-}$ below the one-proton emission threshold. The calculated proton separation energy in ${ {}^{7}\text{Be} }$ is ${ {S}_{ \text{p} }^{ ( \text{th} ) } = 5.613}$ MeV, and agrees well with the experimental value ${ {S}_{ \text{p} }^{ ( \text{exp} ) } = 5.606}$ MeV. The ${ {5/2}_{1}^{-} }$ and ${ {5/2}_{2}^{-} }$ resonances above the one-proton decay threshold should be seen as peaks in $M1$ and $E2$ transitions.

\begin{figure}[htb]
	\includegraphics[width=0.85\linewidth]{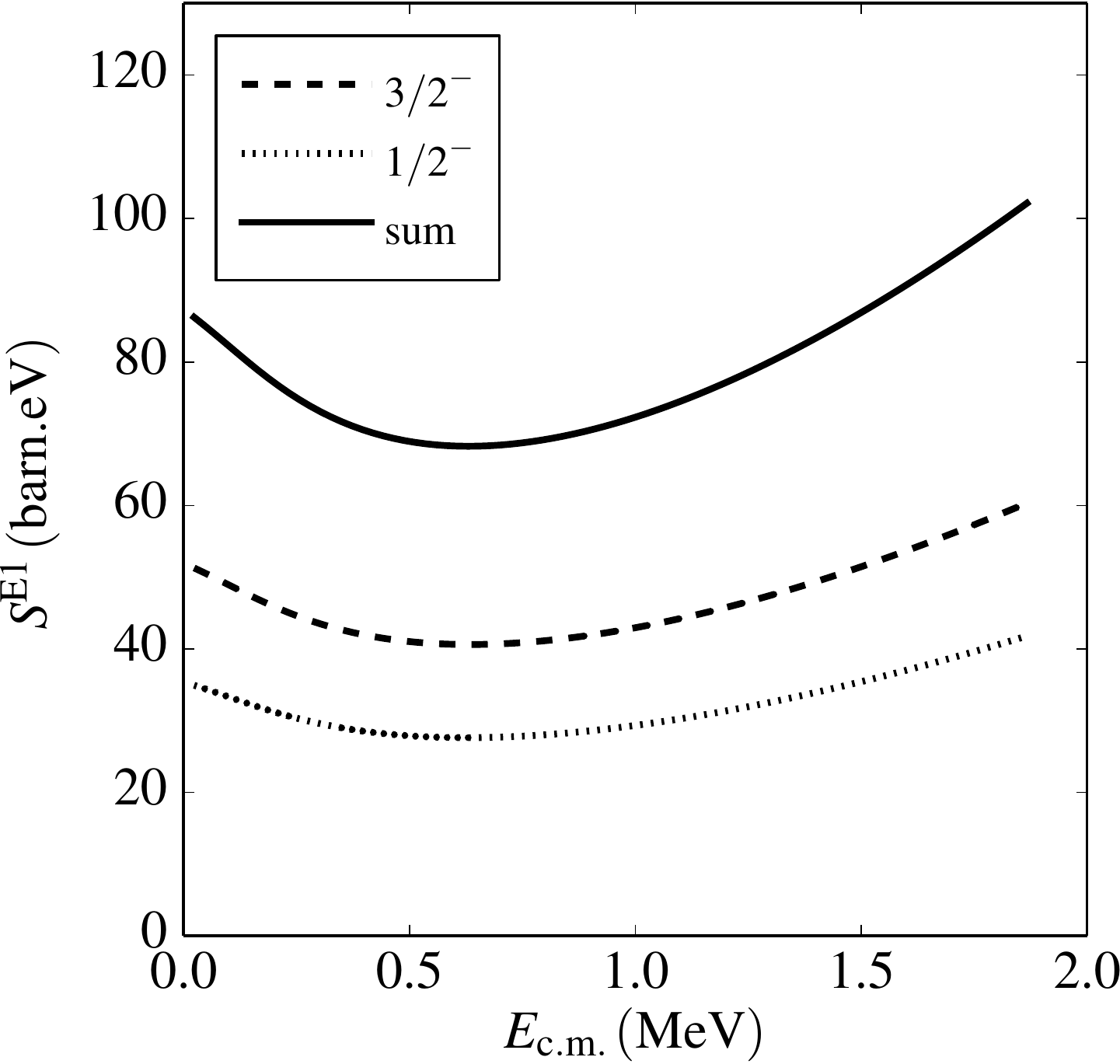}
	\caption{Plot of the $E1$ astrophysical factor for the $^6\text{Li}(p,\gamma)^7\text{Be}$ reaction. The solid line represents the exact, fully antisymmetrized GSM-CC calculation for the capture to both the ground state $J^{\pi}={3/2}_1^-$ and the first excited state $J^{\pi}={1/2}_1^-$ of ${ {}^{7}\text{Be}}$. The dashed and dotted lines show separate contributions for the capture to the ground state and the first excited state, respectively.}
	\label{fig-2}
\end{figure}

\begin{figure}[htb]	
	\includegraphics[width=0.85\linewidth]{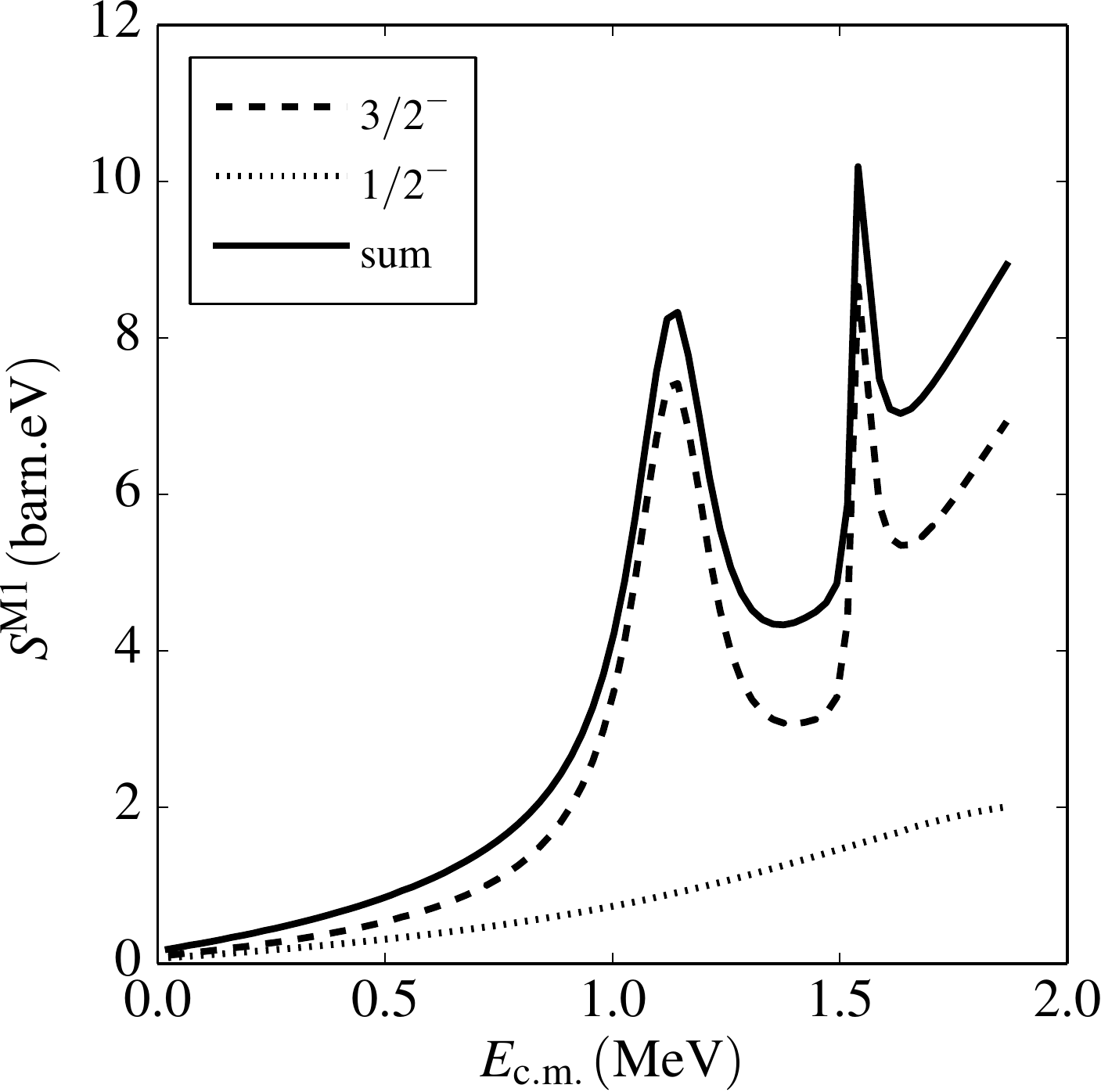}
	\caption{The same as in Fig.\ref{fig-2} but for the $M1$ transitions. The two peaks correspond to the ${5/2}_{1}^-$ and ${5/2}_{2}^-$ resonances of $^7\text{Be}$.}
	\label{fig-3}
\end{figure}

\begin{figure}[htb]
	\includegraphics[width=0.85\linewidth]{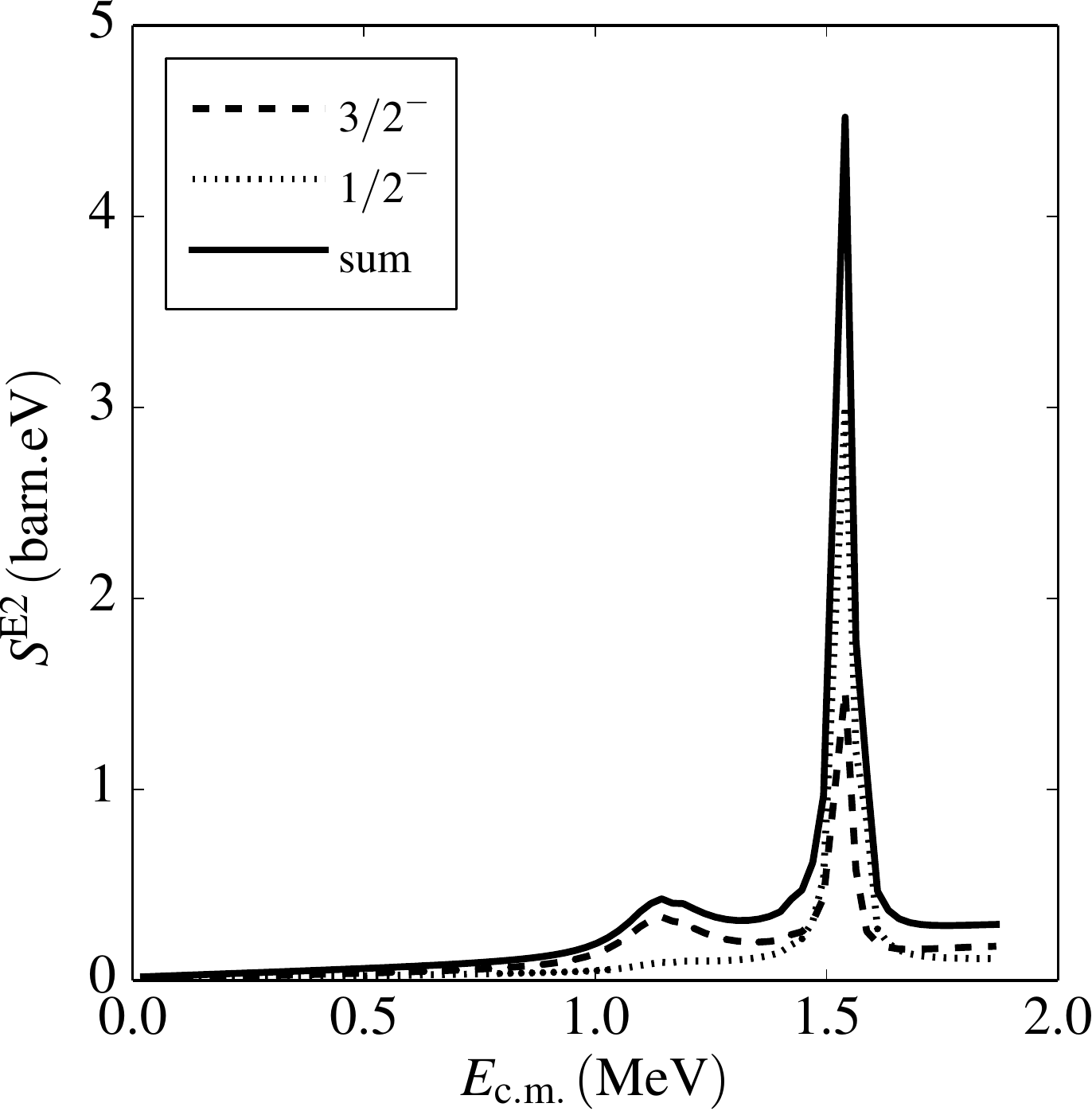}
	\caption{The same as in Fig.\ref{fig-2} but for the $E2$ transitions. The two peaks correspond to the ${5/2}_{1}^-$ and ${5/2}_{2}^-$ resonances of $^7\text{Be}$.}
		\label{fig-4}
\end{figure}

All relevant $E1$, $M1$, $E2$ transitions from the initial continuum states ($J_i = {3/2}^-, {1/2}^-, {7/2}^-, {5/2}^-$) in $^7$Be to the final bound states ($J_f = {3/2}_1^-, {1/2}_1^-$) have been included. Figs.~\ref{fig-2}-\ref{fig-4} show the separate contributions to the total $S$-factor in $^6\text{Li}(p,\gamma)^7\text{Be}$ reaction: $S^{\rm E1}$ for $E1$ transitions (Fig.~\ref{fig-2}), $S^{\rm M1}$ for $M1$ transitions (Fig.~\ref{fig-3}), and $S^{\rm E2}$ for $E2$ transitions (Fig.~\ref{fig-4}). The solid lines in Figs.~\ref{fig-2}-\ref{fig-4} show results of the fully antisymmetrized GSM-CC calculations for the capture to both the ground and the first excited states of $^7$Be composite nuclei. The dashed and dotted lines in these figures correspond to GSM-CC calculations for capture to the ground and first excited states in $^7$Be, respectively.

As compared to $M1$ and $E2$ transitions, the $E1$ transitions contribute most to the total astrophysical $S$ factor. This is consistent with results of previous studies~\cite{Barker80,Arai02} which found that the $E1$ multipolarity and $l=0, 2$ incoming partial waves dominate in the gamma-ray transitions. There is no resonant contribution in $E1$ transitions. The contribution of the capture to the first excited state of ${^7}\text{Be}$ increases the value of $S^{\rm E1}$ factor by $\sim 40 \%$ (see Fig. \ref{fig-2}). 

Although the $S^{\rm M1}$ factor is negligible at low energies, its contribution increases fast in the region of ${5/2}_1^-$ and ${5/2}_2^-$ resonances. In this range of excitation energies, the $M1$ contribution of the capture to the ground state strongly dominates over the contribution of the first excited state (see Fig. \ref{fig-3}). For both $M1$ and $E2$ transitions, one can see ${ {5/2}_{1}^{-} }$ and ${ {5/2}_{2}^{-} }$ resonances of ${ {}^{7}\text{Be} }$ at ${ {E}_{ \text{c.m.} } = 1.126 }$ MeV and ${ {E}_{ \text{c.m.} } = 1.543}$ MeV, respectively. These resonances are observed experimentally at ${ {E}_{ \text{c.m.} } = 1.1242}$ MeV and ${ {E}_{ \text{c.m.} } = 1.604}$ MeV, respectively.

The $E2$ transitions contribute little to the total $S$ factor. In general, $S^{\rm E2}$ is $\sim$10$^{3}$ smaller than $S^{\rm E1}$. Only in the narrow range of excitation energies around the ${5/2}_2^-$ resonance, the contribution of $E2$ transitions become sizable (see Fig. \ref{fig-4}).

\begin{figure}[htb]
	\includegraphics[width=0.85\linewidth]{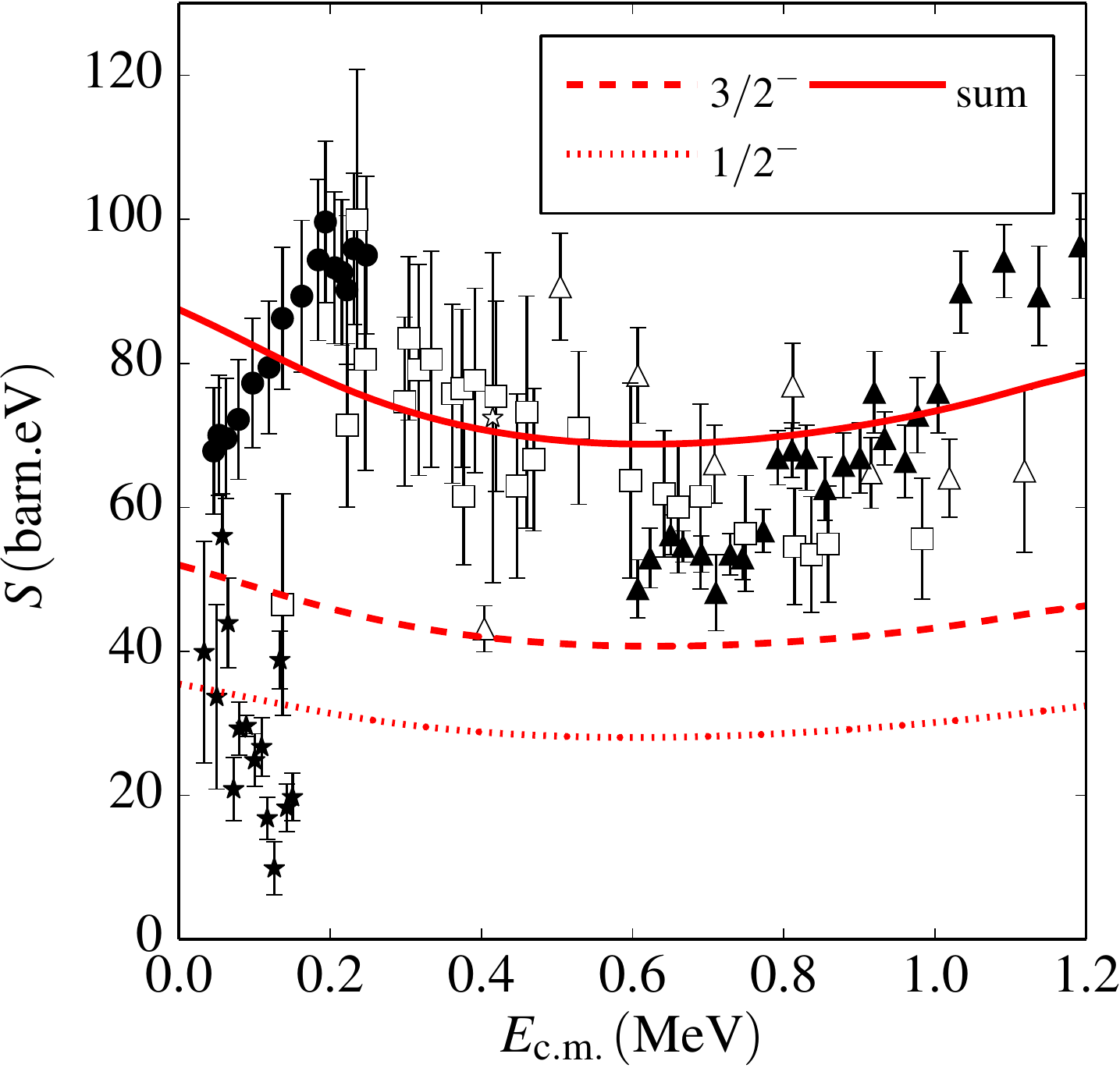}
	\caption{(Color online) Plot of the total astrophysical factor for the $^6\text{Li}(p,\gamma)^7\text{Be}$ reaction. Data are taken from Refs.~\cite{Switkowski79} (open squares), ~\cite{He13} (filled circles), ~\cite{Paradellis99} (filled triangles), ~\cite{Ostojic83} (open triangles), ~\cite{Bashkin55} (open stars), and~\cite{Bruss92} (filled stars-- only ${1/2}^-$ contribution). The solid line represents the exact, fully antisymmetrized GSM-CC calculation for the capture to both the ground state $J^{\pi}=3/2_1^-$ and the first excited state $J^{\pi}=1/2_1^-$ of ${ {}^{7}\text{Be}}$. GSM-CC calculations of separate contributions from the capture to either the ground state or the first excited state of ${ {}^{7}\text{Be}}$ are shown with the dashed and dotted lines, respectively.}
	\label{fig-5}
\end{figure}

The calculated total astrophysical $S$ factor by GSM-CC is compared with the experimental data~\cite{He13,Switkowski79,Paradellis99,Bruss92,Ostojic83,Bashkin55} in Fig.~\ref{fig-5}. Switkowski {\em et al.}~\cite{Switkowski79} (open squares in Fig.~\ref{fig-5}) measured $^6$Li$(p,\gamma)$$^7$Be cross section over a wide energy range of astrophysical interest. These data are well described by the GSM-CC calculations except for the lowest experimental point at $E_\text{c.m.}=$ 140 keV. 

Bruss {\em et al.} \cite{Bruss92} reported the contribution from the capture to the first excited state $J^{\pi}=1/2_1^-$ of ${ {}^{7}\text{Be}}$ (filled stars in Fig.~\ref{fig-5}). A strong decrease of this contribution with increasing c.m. energy is not seen in GSM-CC calculations which, on the contrary, predict a weak dependence on energy of this contribution to the total astrophysical factor $S(E_{\rm c.m.})$.

Recently, He {\em et al.} \cite{He13} (filled circles in Fig.~\ref{fig-5}) reported a sudden drop of the total astrophysical factor $S(E_{\rm c.m.})$ at low energies and predicted a new positive parity resonance, ${1/2}^+$ or ${3/2}^+$, at $E_\text{c.m.}\simeq 0.195$ MeV. We do not confirm this experimental finding in our calculations.

The energy dependence of the astrophysical $S$ factors has been studied by Prior {\em et al.}~\cite{Prior04} who showed that $S(E_{\rm c.m.})$ has a negative slope towards low energies, while an earlier measurement indicated a positive slope~\cite{Cecil92}. In our studies, the slope of $S(E_{\rm c.m.})$ is negative, and $S_{\rm GSM-CC}(0)$=88.34 b$\cdot$eV is close to the accepted experimental value $S_{\rm exp}(0)=$79$\pm$18 b$\cdot$eV.
GSM-CC results agree also qualitatively with other theoretical studies of this reaction~\cite{Barker80,Arai02,Huang10}, but predict a lower value for $S(0)$. 
\begin{figure}[htb]
	\includegraphics[width=0.85\linewidth]{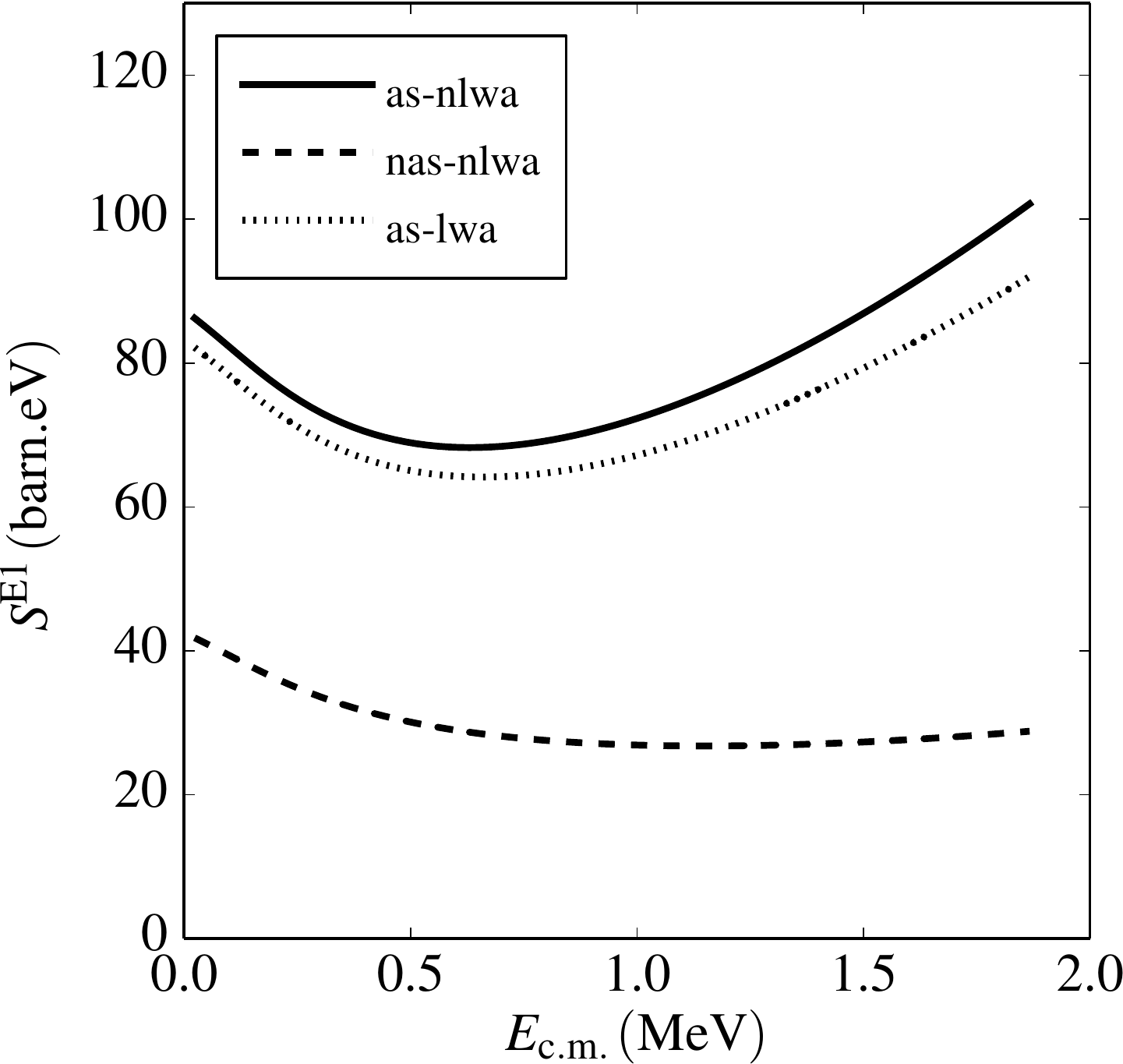}
	\caption{Plot of the $E1$ astrophysical factor for the ${ {}^{6}\text{Li} ( p , \gamma ) {}^{7}\text{Be} }$ reaction. The solid line represents the exact, fully antisymmetrized GSM-CC calculation. The calculations in the long wavelength approximation are represented by the dashed and dotted lines in the fully antisymmetrized and non-antisymmetrized cases, respectively.}
	\label{fig-6}
\end{figure}

\begin{figure}[htb]	
	\includegraphics[width=0.85\linewidth]{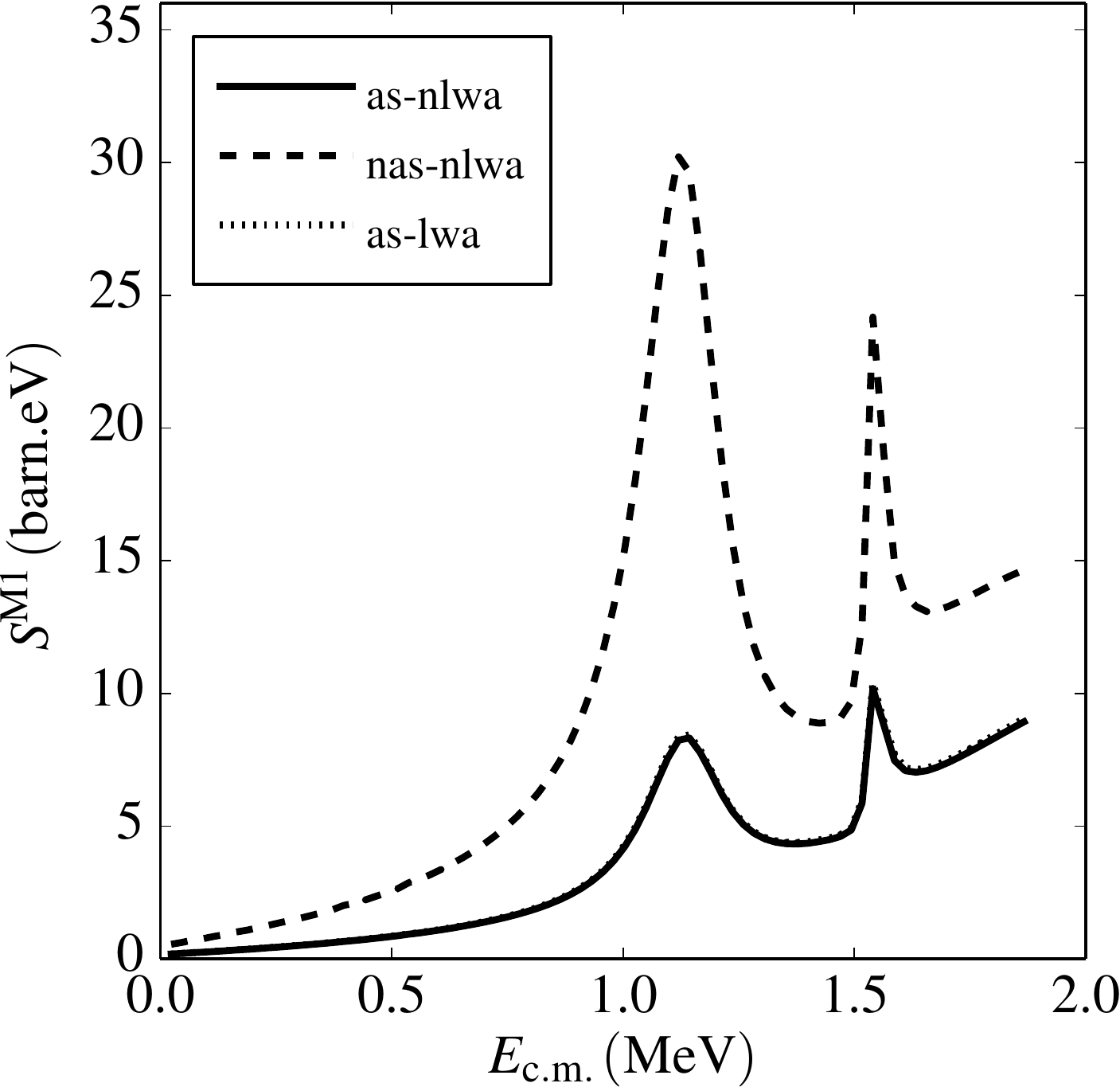}
	\caption{The same as in Fig.\ref{fig-6} but for the $M1$ transitions. The two peaks correspond to the ${ {5/2}_{1}^{-} }$ and ${ {5/2}_{2}^{-} }$ resonances of ${ {}^{7}\text{Be} }$.}
	\label{fig-7}
\end{figure}

\begin{figure}[htb]
	\includegraphics[width=0.85\linewidth]{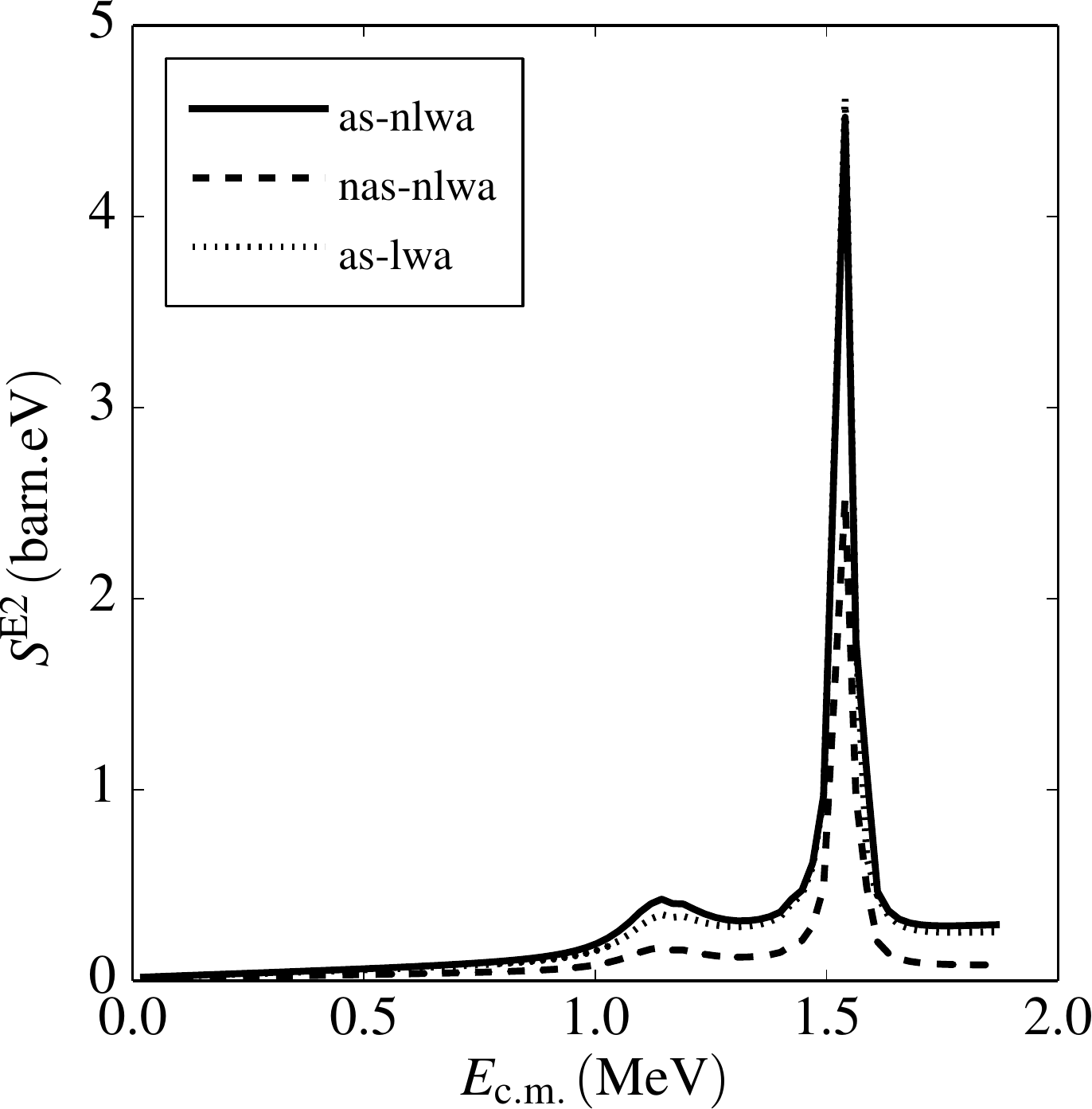}
	\caption{The same as in Fig.\ref{fig-6} but for the $E2$ transitions. The two peaks correspond to the ${ {5/2}_{1}^{-} }$ and ${ {5/2}_{2}^{-} }$ resonances of ${ {}^{7}\text{Be} }$.}
		\label{fig-8}
\end{figure}

The long-wavelength approximation simplifies the calculation of matrix elements of the electromagnetic transitions. The quality of this approximation and the role of the antisymmetry of initial and final states is tested in Figs. \ref{fig-6}-\ref{fig-8}. For $E1$, both the long-wavelength approximation and the absence of antisymmetry in the channel state ${ \ket{ r , c } }$, i.e.  $\ket{ r , c } \equiv \ket{r} \otimes \ket{c}$, decrease the $E1$ contribution to the total astrophysical $S$ factor (see Fig.~\ref{fig-6}). However, whereas the long-wavelength approximation reduces the $S$ factor by $\sim 5 \%$, the absence of the antisymmetry in the calculation of the matrix elements of the electromagnetic operators reduces it by almost $\sim 50 \%$. 
For $M1$ and $E2$ transitions, the long-wavelength approximation nearly does not change $S^{\rm M1}$ and $S^{\rm E2}$, while the lack of antisymmetry of initial and final states increases the value of $S^{\rm M1}$ by a factor ${ \sim 4 }$ at the first resonance peak and a factor ${ \sim 2 }$ at the second resonance peak (see Fig.~\ref{fig-7}). The antisymmetrization is significant for $S^{\rm E2}$ only at the resonance peak (see Fig.~\ref{fig-8}).

\subsection{Results for the $^6\text{Li}(n,\gamma)^7\text{Li\emph{}}$ reaction}
\label{sec3b}

In this section, we shall discuss the mirror reaction of the radiative proton capture reaction $^6$Li$(p,\gamma)$$^7$Be. Both reactions are described in the same valence space with the shells in the non-resonant continuum taken along the same discretized contours: $\mathcal{L}^+_{s_{1/2}}$, $\mathcal{L}^+_{p_{1/2}}$, $\mathcal{L}^+_{p_{3/2}}$, $\mathcal{L}^+_{d_{3/2}}$, and $\mathcal{L}^+_{d_{5/2}}$. The WS potential of $^4$He core in GSM and GSM-CC studies of $^6$Li, $^7$Li, and$^6\text{Li}(n,\gamma)^7\text{Li\emph{}}$ is given in Table~\ref{para-3}. The radius of the Coulomb potential is $r_{\text{Coul}}$=2.5 fm. Parameters of the FHT Hamiltonian are given in Table~\ref{para-4}.

\begin{table}[ht!]
\caption{Parameters of the WS potential of $^4$He core used in the description of $^6$Li, $^7$Li nuclei, and $^6\text{Li}(n,\gamma)^7\text{Li\emph{}}$ reaction.\label{para-3}}
\begin{ruledtabular}
\begin{tabular}{ccc}
  Parameter & Protons & Neutrons \\
  \hline
  $a$ & 0.65 fm & 0.65 fm \\
  $R_0$ & 2.0 fm & 2.0 fm \\
  $V_\text{o}$ & 53.122 MeV & 46.468 MeV \\
  $V_\text{so}(l=1)$ & 3.208 MeV & 6.27 MeV \\
  $V_\text{so}(l=2)$ & 3.208 MeV & 6.27 MeV \\
\end{tabular}
\end{ruledtabular}
\end{table}

\begin{table}[ht!]
\caption{Parameters of the FHT interaction in GSM and GSM-CC calculations in $^6$Li, $^7$Li, and $^6\text{Li}(n,\gamma)^7\text{Li\emph{}}$. For more details, see the caption of Table~\ref{para-2}.\label{para-4}}
\begin{ruledtabular}
\begin{tabular}{cc}
  Parameter & Value  \\
  \hline
  $\nu^\textsf{C}_{\textsf{t,t}}$  & 16.868 MeV\\
  $\nu^\textsf{C}_{\textsf{s,t}}$  & -8.587 MeV\\
  $\nu^\textsf{C}_{\textsf{s,s}}$  & -14.176 MeV\\
  $\nu^\textsf{C}_{\textsf{t,s}}$  & -7.683 MeV\\
  $\nu^\textsf{SO}_{\textsf{t,t}}$  & -1181.015 MeV\\
  $\nu^\textsf{SO}_{\textsf{s,t}}$  & 0 MeV\\
  $\nu^\textsf{T}_{\textsf{t,t}}$  & 17.852 MeV fm$^{-2}$\\
  $\nu^\textsf{T}_{\textsf{s,t}}$  & -1.241 MeV fm$^{-2}$\\
\end{tabular}
\end{ruledtabular}
\end{table}

\subsubsection{Spectrum of $^7$Li}

For the parameters of the GSM Hamiltonian given in Tables \ref{para-3} and \ref{para-4}, the ground state of $^6$Li is bound by 3.70016 MeV with respect to $^4$He.
Reaction channels are generated by coupling the GSM states $J^{\pi}_{\rm targ}=1^+, 3^+_1, 0^+_1$, and $2^+_1$ of $^6$Li with the neutron in partial waves: $s_{1/2}$, $p_{1/2}$, $p_{3/2}$, $d_{3/2}$ and $d_{5/2}$. The composite states of $^7$Li $([^6\text{Li}(J_{\text{targ}}^\pi)\otimes n(l,j)]^{J_{\text{targ}}^\pi})$ are ${3/2}_1^-$, ${1/2}_1^-$ bound states, and ${7/2}_1^-$, ${5/2}_1^-$, ${5/2}_2^-$ resonances.

\begin{figure}[htb]
\includegraphics[width=0.85\linewidth]{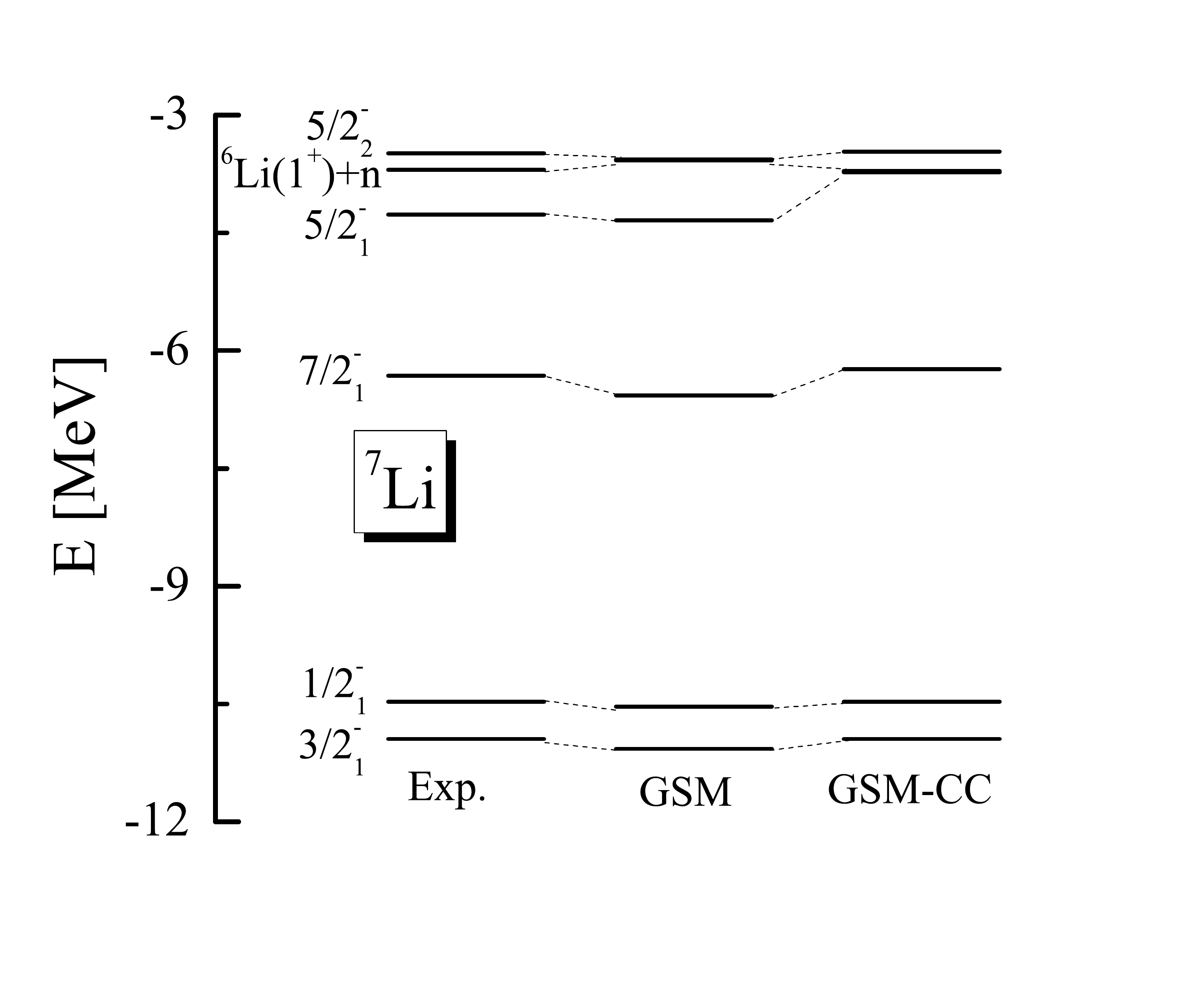}
\caption{Energy spectrum in GSM and GSM-CC calculation for $^7$Li. Energies are given relative to the energy of $^4$He core.}
\label{fig-9}
\end{figure}

The energy spectrum of states in $^7$Li and the neutron separation energy are compared in Fig.~\ref{fig-9} with the experimental data~\cite{nndc}. In GSM-CC calculations, the two-body part of ${\hat H}$ for ${J^\pi=3/2}_1^-$, ${1/2}_1^-$ states has been rescaled by the small multiplicative factors: $c(\frac{3}{2}^-)=1.02$, $c(\frac{1}{2}^-)=1.047$ to correct for the omission of non-resonant channels that are built by coupling the continuum states of $^6$Li with the neutron in different partial waves. 

One can see in Fig.~\ref{fig-9}, that the experimental data are well reproduced by the GSM and GSM-CC calculations.
${7/2}^-_1$ and ${5/2}_1^-$ states, which lie in-between the $[{^4}\text{He} - {^3}\text{He}]$ are bound in our GSM-CC calculation because the $[^4\text{He}(J_{\rho}^\pi)\otimes ^3\text{H}(J_{\rho}^\pi)]^{J_f^\pi}$ reaction channel is absent in the basis.
The resonance ${5/2}^-_2$ ($\Gamma^{\rm exp}\simeq 89$ keV), lying above the $[{^4}\text{He} - {^3}\text{He}]$ and  
$[{^6}\text{Li} - \text{n}]$ decay thresholds, is also a resonance in our studies. The calculated width for this state $\Gamma^{\rm th}\simeq 38$ keV is smaller than found experimentally, partially due to the missing $[^4\text{He}(J_{\rho}^\pi)\otimes ^3\text{H}(J_{\rho}^\pi)]^{J_f^\pi}$ reaction channel.

\subsubsection{Cross section for $^6$Li$(n,\gamma)$$^7$Li reaction}

The experimental neutron separation energy in $^7\text{Li}$ is $S_{\rm n} = 7.249$ MeV. This nucleus has four bound states $J^{\pi}={3/2}_1^-$, ${1/2}_1^-$, ${7/2}_1^-$, and ${5/2}_1^-$ below the neutron emission threshold. The calculated neutron separation energy is $S_{\text{n}}^{\text{th}}=7.242$ MeV, in good agreement with the experimental data. The ${5/2}_2^-$ resonance above the one-neutron decay threshold can be seen as the peak both in $M1$ and $E2$ transitions.

\begin{figure}[htb]
	\includegraphics[width=0.85\linewidth]{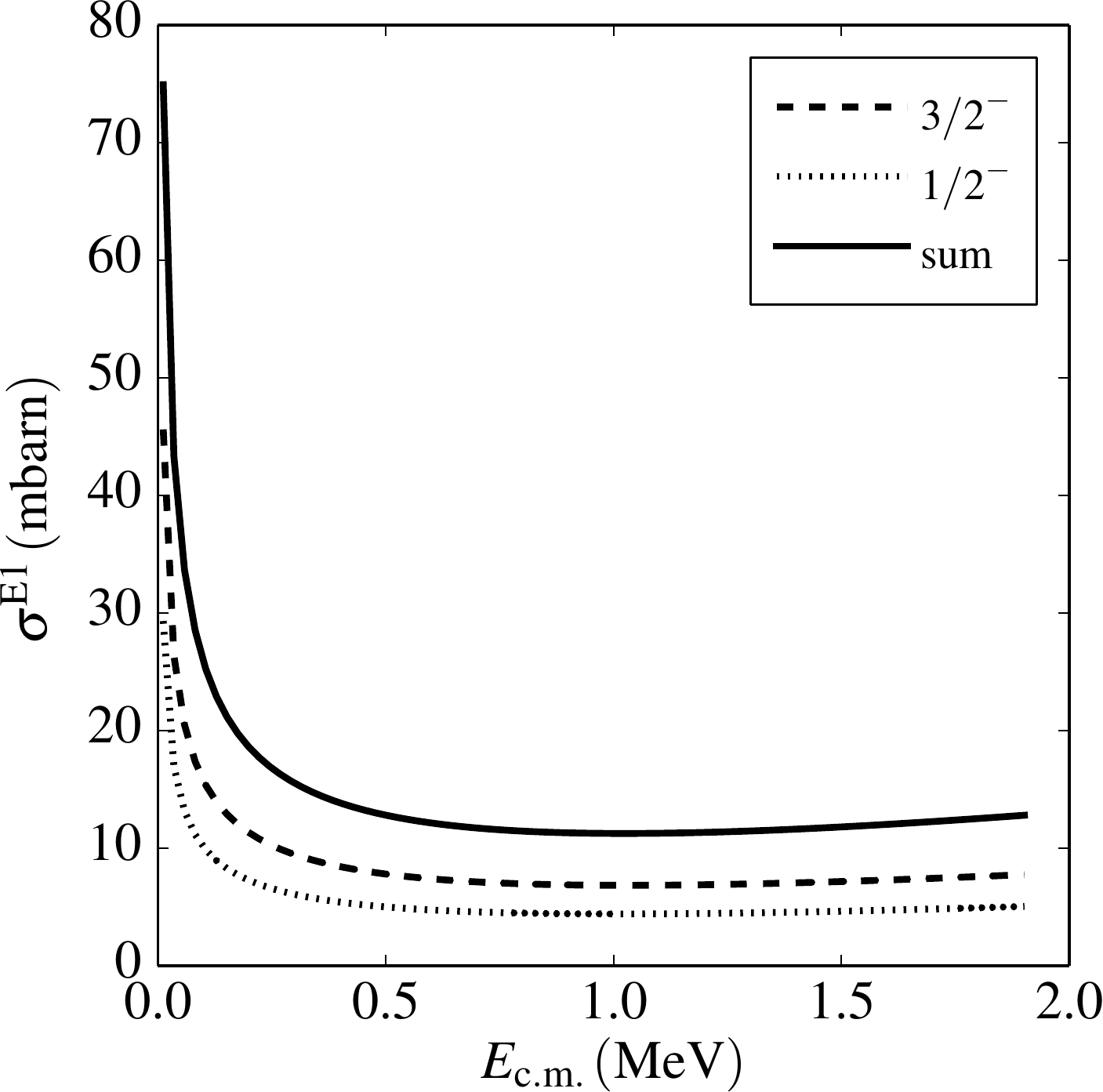}
	\caption{The same as in Fig.\ref{fig-2} but for the ${ {}^{6}\text{Li} ( n , \gamma ) {}^{7}\text{Li} }$ reaction.}
	\label{fig-10}
\end{figure}

\begin{figure}[htb]	
	\includegraphics[width=0.85\linewidth]{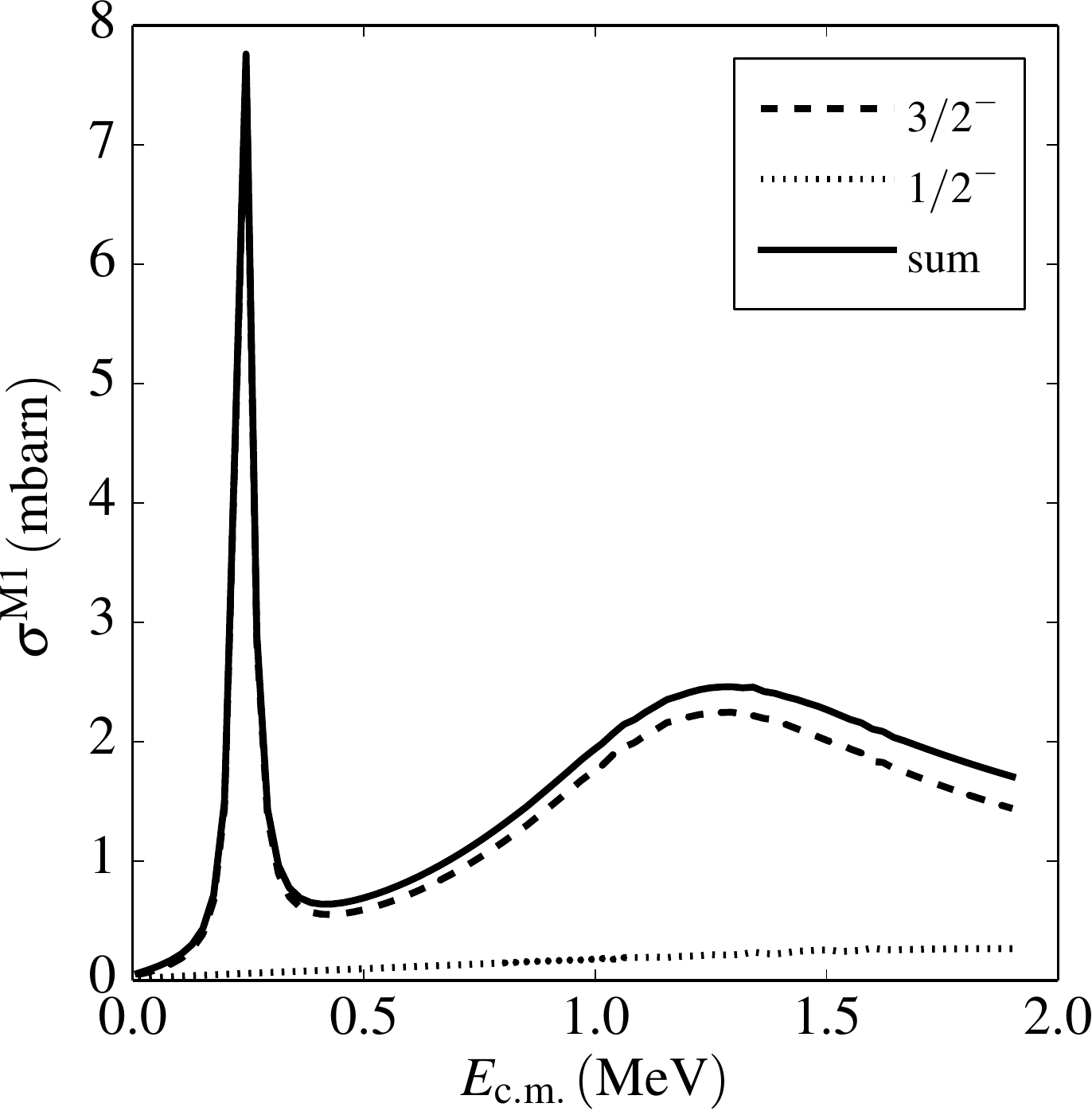}
	\caption{The same as in Fig.\ref{fig-2} but for the $M1$ transitions in the ${ {}^{6}\text{Li} ( n , \gamma ) {}^{7}\text{Li} }$ reaction. The peak corresponds to the ${5/2}_{2}^-$ resonance of $^7\text{Li}$.}
	\label{fig-11}
\end{figure}

\begin{figure}[htb]
	\includegraphics[width=0.85\linewidth]{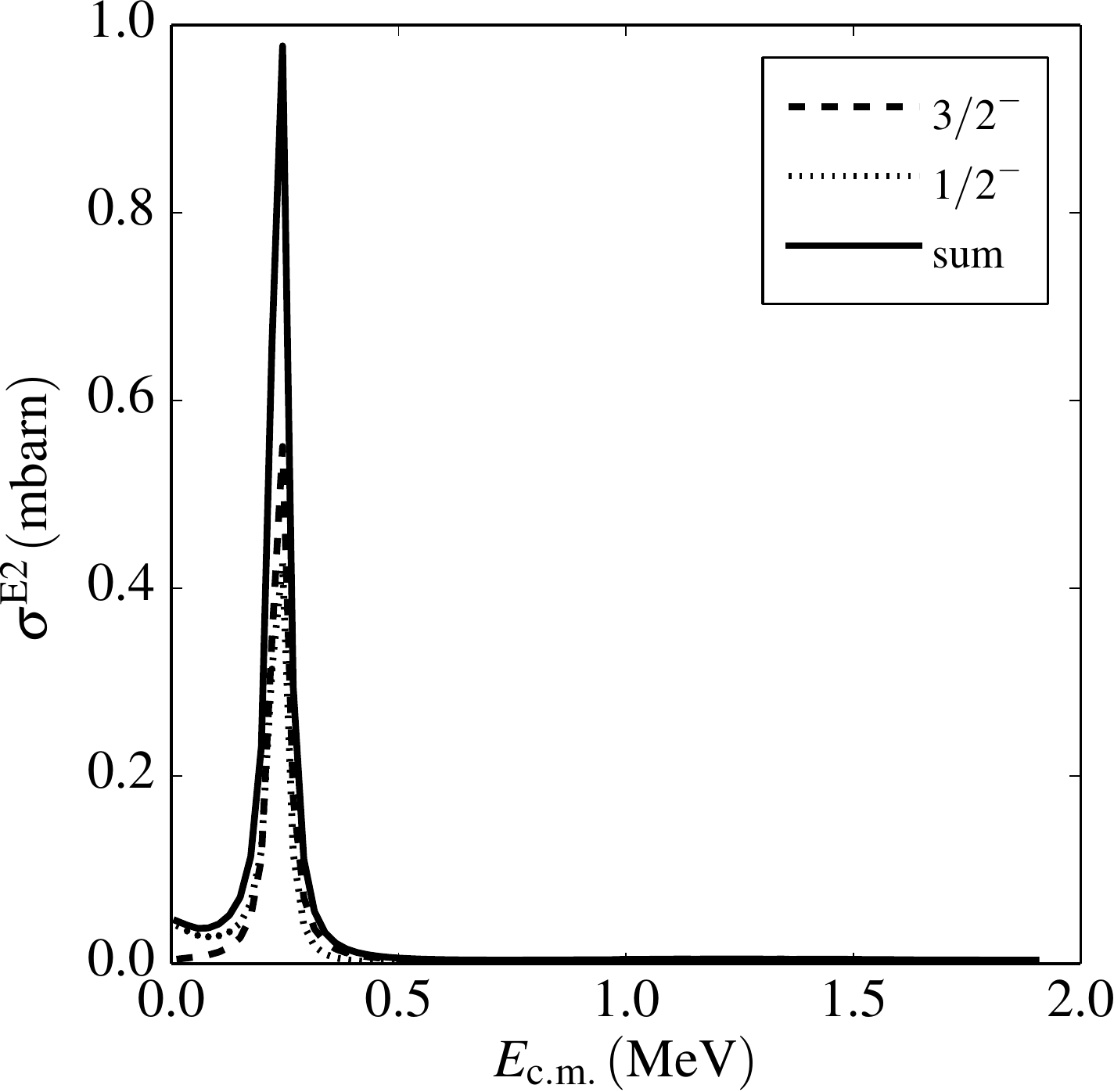}
	\caption{The same as in Fig.\ref{fig-2} but for the $E2$ transitions in the ${ {}^{6}\text{Li} ( n , \gamma ) {}^{7}\text{Li} }$ reaction. The peak corresponds to the ${5/2}_{2}^-$ resonance of $^7\text{Li}$.}
		\label{fig-12}
\end{figure}

Figs.~\ref{fig-10}-\ref{fig-12} show separate contributions from $E1$, $M1$ and $E2$ transitions to the total cross section of the reaction $^6\text{Li}(n,\gamma)^7\text{Li}$. All relevant transitions from the initial continuum states ($J_i = {3/2}^-, {1/2}^-, {7/2}^-, {5/2}^-$) to the final bound states $J_f = {3/2}_1^-, {1/2}_1^-$ are included. The solid lines in Figs.~\ref{fig-10}-\ref{fig-12} show results of the fully antisymmetrized GSM-CC calculations for the radiative neutron capture to both the ground state and the first excited state of $^7$Li. The dashed and dotted lines in these figures show contributions from the capture to the ground state and the first excited state, respectively.

From Figs.~\ref{fig-10}-\ref{fig-12}, we can see that $E1$ transitions contribute most to the total neutron radiative capture cross section. There is no resonant contribution in $E1$ transitions. The capture to the first excited state increases the $E1$ part of the total cross section by $\sim 40 \%$.  

In $M1$ and $E2$ transitions, the resonant contributions are not negligible. One can see a fast growth of the cross-section around the ${5/2}_2^-$ resonance (see Fig. \ref{fig-12}). In this resonant $M1$ contribution, the radiative neutron capture to the first excited state is negligible. The ${ {5/2}_{2}^{-} }$ resonance can be seen in both $M1$ and $E2$ transitions. The peak of the resonance is at ${ {E}_{ \text{c.m.} } = 0.2383 }$ MeV in GSM-CC calculations, in agreement with the data (${ {E}_{ \text{c.m.} } = 0.2096}$ MeV).

$E2$ provides the least contribution to the total cross section. Except for the narrow region excitation energies around the ${5/2}_2^-$ resonance, $\sigma^{\rm E2}$ is smaller by a factor $\sim$10$^{3}$ than $\sigma^{\rm E1}$ (see Fig. \ref{fig-12}).

\begin{figure}[htb]
	\includegraphics[width=0.85\linewidth]{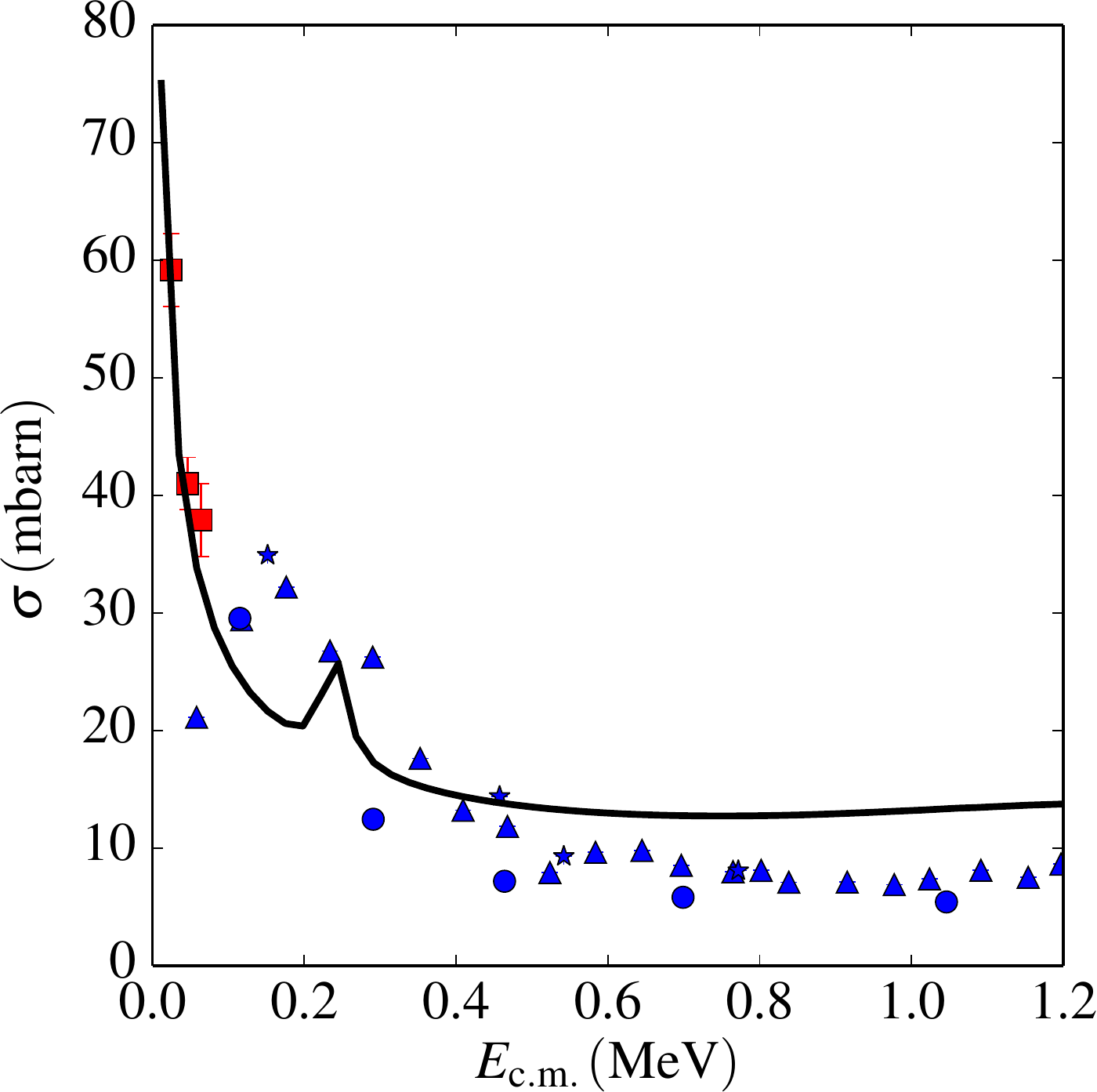}
	\caption{Plot of the total cross section for the reaction ${ {}^{6}\text{Li} ( n , \gamma ) {}^{7}\text{Li} }$. The data are taken from Ref.~\cite{Ohsaki00} (filled squares), Ref.~\cite{Karataglisis89} (filled triangles), Ref.~\cite{Bramblett73} (filled circles), and Ref.~\cite{Green64} (filled stars). The solid line shows results of the fully antisymmetrized GSM-CC calculation of the neutron radiative capture cross section to both the ground state $J^{\pi}=3/2_1^-$ and the first excited state $J^{\pi}=1/2_1^-$ of ${ {}^{7}\text{Li}}$. }
	\label{fig-13}
\end{figure}

The total cross sections for the ${ {}^{6}\text{Li} ( n , \gamma ) {}^{7}\text{Li} }$ reaction is compared with the experimental data~\cite{Ohsaki00,Karataglisis89,Bramblett73,Green64} in Fig.~\ref{fig-13}. The three data points at stellar energies \cite{Ohsaki00} obtained by the direct measurement of the cross section in this reaction, are well reproduced by our calculations. At higher energies (${ {E}_{ \text{c.m.} } }>0.5 \,\text{MeV}$), the calculated cross-sections overshoot the experimental data~\cite{Karataglisis89,Bramblett73,Green64} by $\sim 20 \%$. These data have been extracted from the photodisintegration cross sections of $^7$Li into the $n - ^6$Li channels.

The neutron radiative capture cross section at very low energies can be extracted using the expansion:
\begin{equation}
	\sigma ( {E}_{ \text{c.m.} } ) = \frac{8.12828}{ \sqrt{ {E}_{ \text{c.m.} } } } - 0.496429 + 2.78499 \sqrt{ {E}_{ \text{c.m.} } }
	\label{eq_fit_cross_section}
\end{equation}
Applying this formula to GSM-CC numerical data, one finds: ${ { \sigma }^{ \text{(GSM-CC)} } = 51.41 \, \unitsText{mb} }$ at $E_{\rm c.m.} = 0.025$ eV, in a fair agreement with the experimental data reported in Refs.~\cite{Jarczyk64,Molnar00,Molnar02}, and slightly above the data given in Ref. \cite{Park06,Bartholomew57,Jurney73}.

\begin{figure}[htb]
	\includegraphics[width=0.85\linewidth]{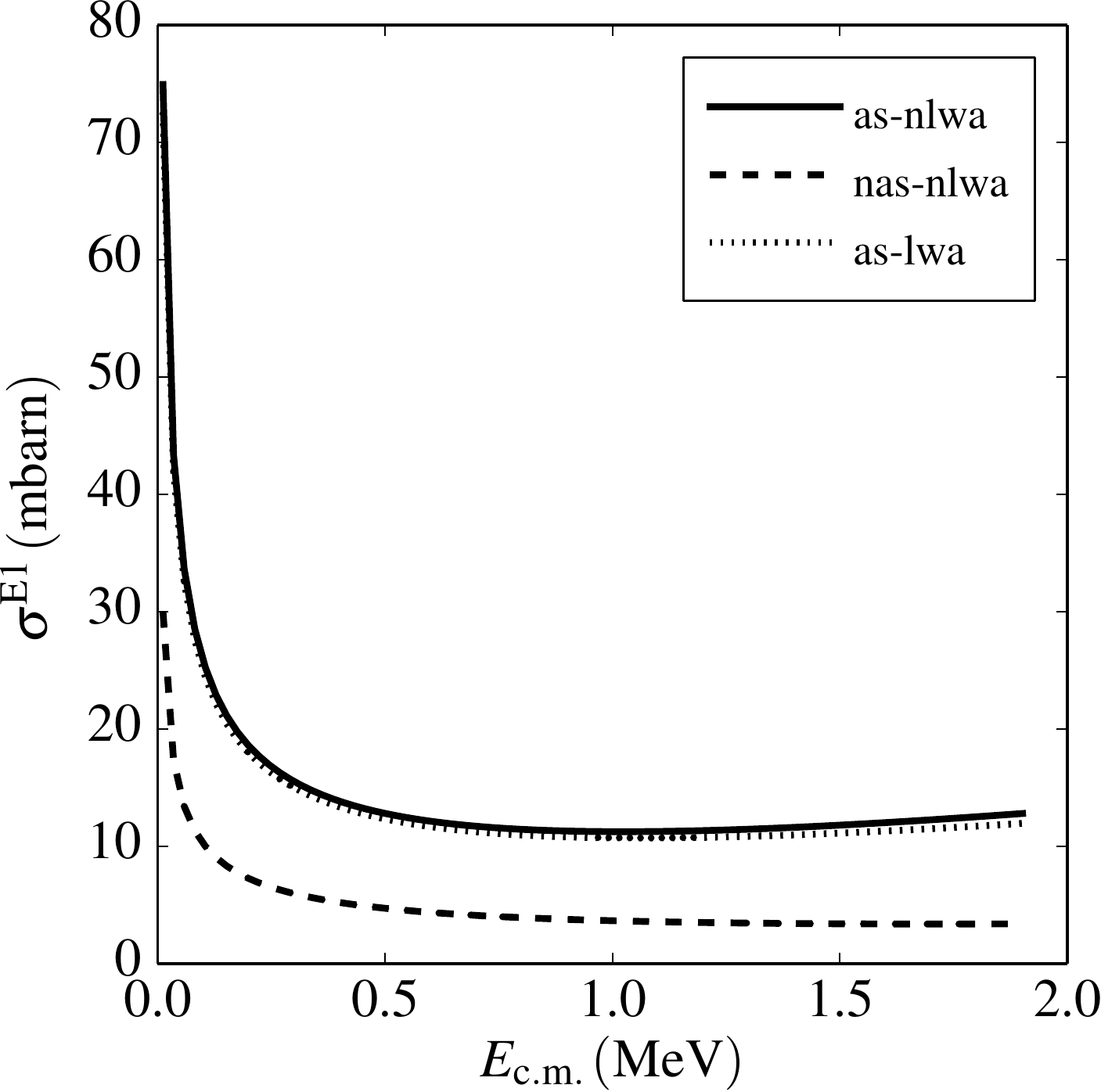}
	\caption{The same as in Fig.\ref{fig-6} but for the $^6\text{Li}(n,\gamma)^7\text{Li}$ reaction.}
	\label{fig-14}
\end{figure}

\begin{figure}[htb]
	\includegraphics[width=0.85\linewidth]{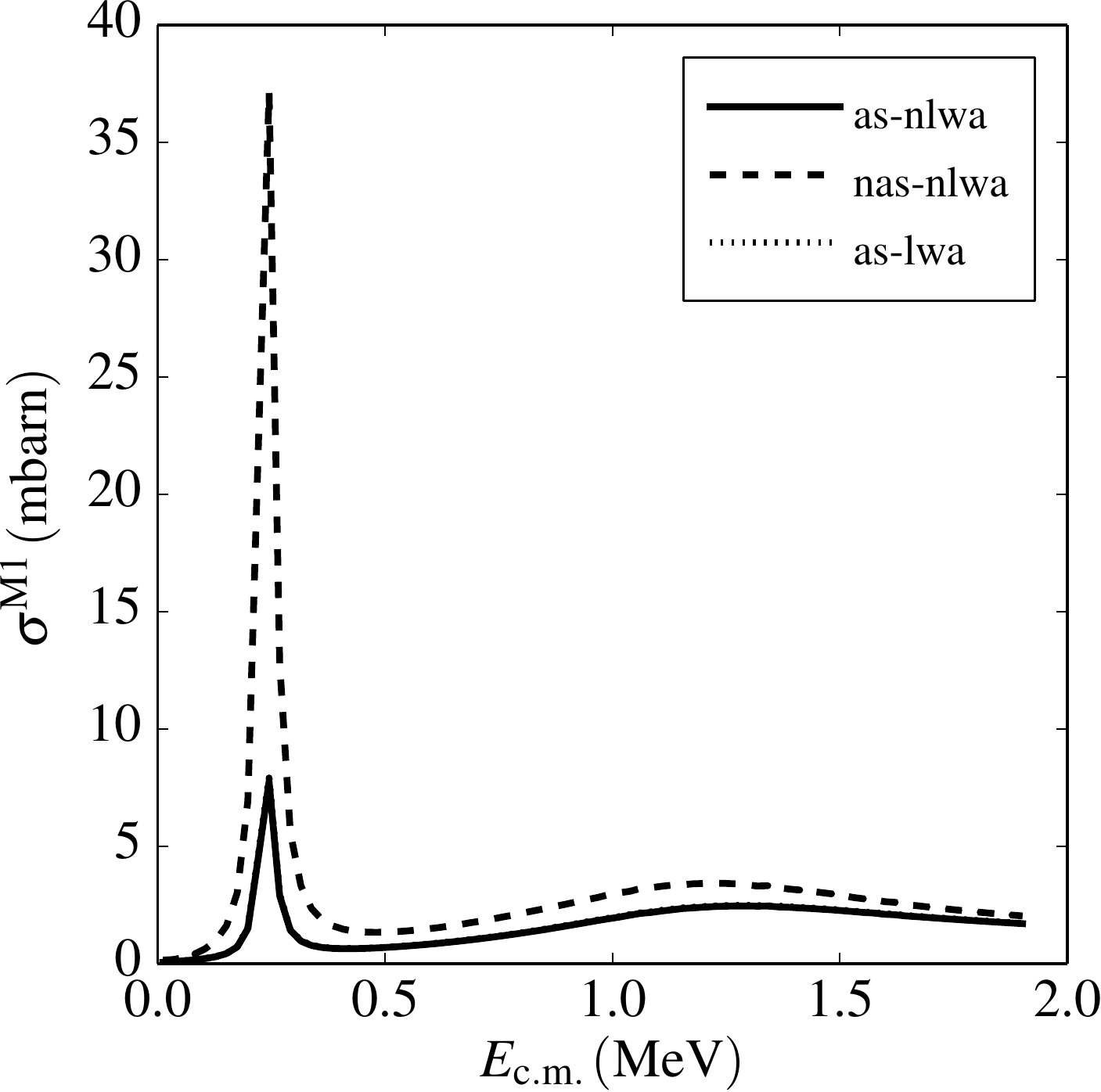}
	\caption{The same as in Fig.\ref{fig-6} but for $M1$ transitions in ${ {}^{6}\text{Li} ( n , \gamma ) {}^{7}\text{Li} }$ reaction. The peak corresponds to the ${5/2}_{2}^-$ resonance in ${ {}^{7}\text{Li} }$.}
	\label{fig-15}
\end{figure}

\begin{figure}[htb]
	\includegraphics[width=0.85\linewidth]{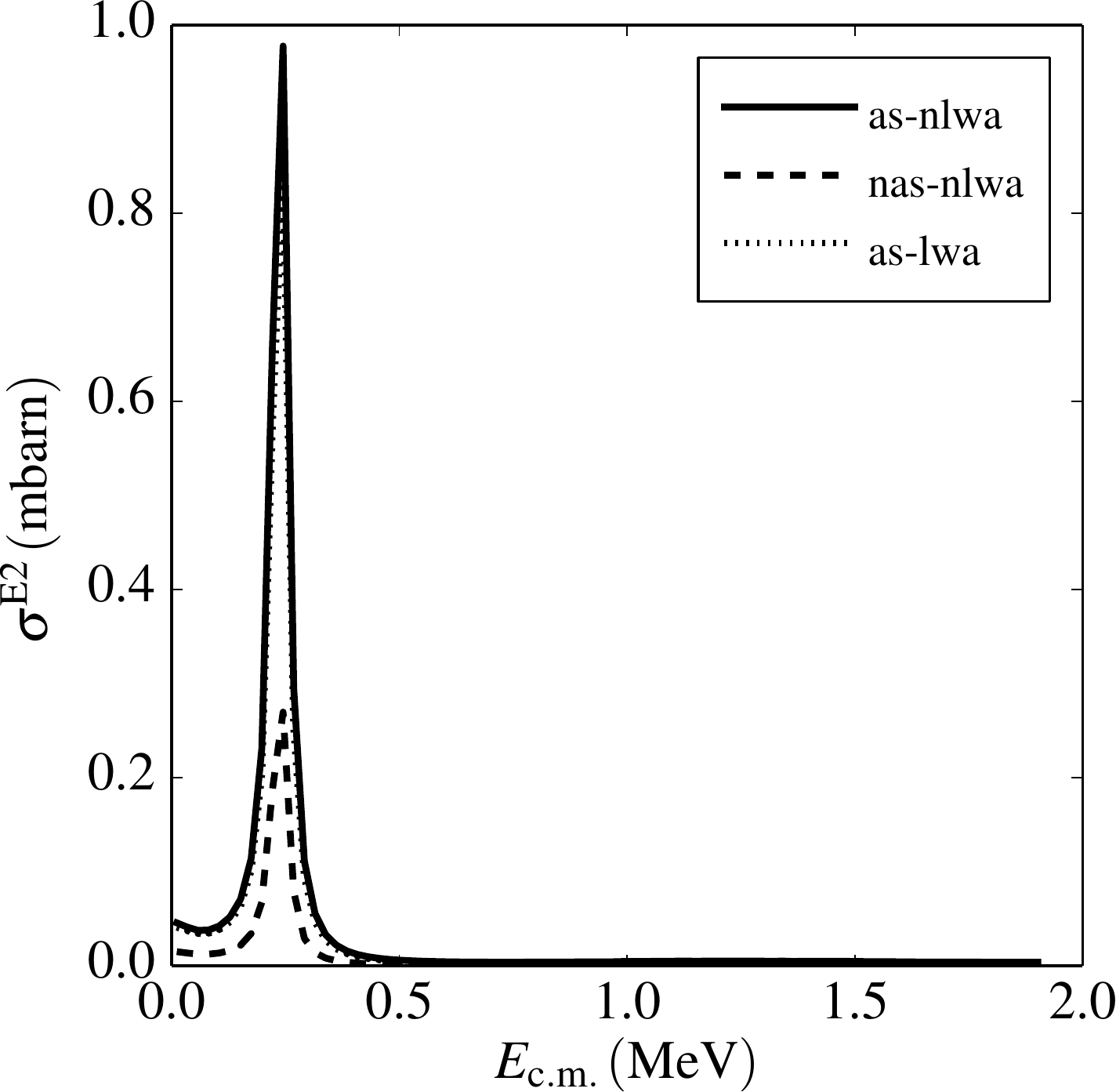}
	\caption{The same as in Fig.\ref{fig-6} but for the $E2$ transitions in ${ {}^{6}\text{Li} ( n , \gamma ) {}^{7}\text{Li} }$ reaction. The peak corresponds to the ${5/2}_{2}^-$ resonance of ${ {}^{7}\text{Li} }$.}
	\label{fig-16}
\end{figure}

The long-wavelength approximation and the role of the antisymmetry of initial and final states in the calculation of matrix elements of the electromagnetic transitions is investigated in Figs.~\ref{fig-14}-\ref{fig-16}. One can see that the long-wavelength approximation in the calculation of $E1$, $M1$ or $E2$ transition matrix elements for this reaction is nearly perfect. On the contrary, the antisymmetry of initial and final states is essential for the calculation of transition matrix elements. It enhances strongly the $E1$ component in the radiative neutron capture cross section, even at low excitation energies (see Fig.~\ref{fig-14}), and diminishes the $M1$ component in this cross sections by a factor $\sim5$ at the ${5/2}_2^-$ resonance peak (see Fig.~\ref{fig-15}). Moreover, the $E2$ contribution to the neutron radiative capture cross section increases by a factor $\sim4$ at this resonance (see Fig.~\ref{fig-16}) if the antisymmetry is carried out exactly.

\section{Conclusions}
\label{sec4}

GSM in the coupled-channel formulation provides the unified description of low-energy nuclear structure and reactions. In the present studies, we have applied the GSM-CC approach to describe proton and neutron radiative capture processes on $^6$Li. These reactions are of interest in nuclear astrophysics, mainly in connection with the problem of puzzled abundances of $^6$Li and $^7$Li isotopes. 
According to standard stellar evolution scenario, the lithium content is considered to be an indicator of the stellar age.  It is easily destroyed at relatively low temperatures in mixing processes between stellar surface and hot internal layers so that at the end of the stellar lifetime the lithium content is believed to be depleted. However, a large spread in lithium abundances have been observed also among the evolved stars, on the red giant branch and the asymptotic giant branch, what is one of the puzzles of modern astrophysics. 

In our studies of $^6\text{Li}(p,\gamma)^7\text{Be}$  and $^6\text{Li}(n,\gamma)^7\text{Li}$ reactions, we used a translationally invariant Hamiltonian with the finite-range FHT interaction between valence nucleons. Parameters of this interaction have been adjusted to reproduce spectra of $^6$Li, $^7$Li, $^7$Be as well as the relevant one-nucleon separation energies. 
As compared to the GSM, the configuration space in the GSM-CC  is limited both by the restriction on a number of excitations into the states of a non-resonant continuum and the absence of reaction channels build from the non-resonant continuum states of $^6$Li. These two approximations in GSM-CC numerical calculations are corrected {\em a posteriori} by the tiny adjustment of channel-channel coupling potentials. The calculated spectra of $^6$Li, $^7$Li, and $^7$Be, as well as the proton (neutron) radiative capture cross section in the $^6\text{Li}(p,\gamma)^7\text{Be}$ ($^6\text{Li}(n,\gamma)^7\text{Li})$ reaction are in good agreement with the data. 

In both reactions, we find that the $s$-wave radiative capture of proton (neutron) to the first excited state $J^{\pi}=1/2_1^+$ of $^7$Be ($^7$Li) is important and increases the total astrophysical $S$-factor by about 40 \%. The $s$-wave capture of nucleons in these mirror reactions is particularly important for the $E1$ component of the $S$-factor. It is also essential for the $E2$ contribution at the $5/2_2^+$ resonance. 

At present, the experimental information about the role of the first excited state in these reactions is restricted to a single measurement \cite{Bruss92} in $^6\text{Li}(p,\gamma)^7\text{Be}$. This is insufficient to test the GSM-CC calculations which yield a correct magnitude of this contribution but its energy dependence is different than reported by Bruss {\em et al.} \cite{Bruss92}.

In our studies we found that the antisymmetry of the initial and final states plays a key role in the calculation of the matrix elements of $E1$ and $M1$ operators at all energies of the astrophysical interest, and matrix elements of $E2$ operator at the position of the $5/2_2^+$ resonance. On the contrary, the long-wavelength approximation in the calculation of radiative capture cross-sections is safe in these mirror reactions. 

In future, the inclusion of non-resonant channels build by the non-resonant scattering states of a target nucleus coupled to the incoming nucleon in different partial waves will further increase the intrinsic consistency of the GSM-CC calculation. The extension of the GSM-CC to describe reactions with composite projectiles ($d$,$\alpha$...) is now in progress and will allow to include the reaction channels like 
$[^4\text{He}(J_{\rho}^\pi)\otimes ^3\text{H}(J_{\rho}^\pi)]^{J_f^\pi}$ which are neglected at present but are important to describe the decay width of $5/2_1^+$, $5/2_2^+$ resonances in $^7$Be and $7/2_1^+$, $5/2_1^+$, and $5/2_2^+$ resonances in $^7$Li.

\section{Acknowledgements}
This material is based upon work supported by the U.S. Department of Energy, Office of Science, Office of Nuclear Physics under award numbers DE-SC0013365 (Michigan State University) and DE-FG02-10ER41700 (French-U.S. Theory Institute for Physics with Exotic Nuclei). One of us (G.X. Dong) would like to thank for the support of the Helmholtz Association (HGF) through the Nuclear Astrophysics Virtual Institute (VH-VI-417).

\end{document}